\documentclass[conference]{IEEEtran}
\IEEEoverridecommandlockouts

\usepackage{cite}
\usepackage{amsmath,amssymb,amsfonts}
\usepackage{algorithmic}
\usepackage{graphicx}
\usepackage{textcomp}
\usepackage{xcolor}
\def\BibTeX{{\rm B\kern-.05em{\sc i\kern-.025em b}\kern-.08em
    T\kern-.1667em\lower.7ex\hbox{E}\kern-.125emX}}

\usepackage[linesnumbered,ruled,vlined]{algorithm2e}
\usepackage{multirow}
\usepackage{pifont}
\usepackage{makecell}
\usepackage{subfigure}

\usepackage{booktabs}
\usepackage{hyperref}

\hypersetup{
	colorlinks=true,
	linkcolor=black,
	filecolor=black,      
	urlcolor=teal,
	citecolor=black,
}

\newcommand{\figsummary}[1]{\textbf{#1}}

\newcommand{\cmark}{\ding{51}}

\begin{document}

\title{DTC: Real-Time and Accurate Distributed Triangle Counting in Fully Dynamic Graph Streams}

\author{\IEEEauthorblockN{Wei Xuan\IEEEauthorrefmark{1},
Yan Liang\IEEEauthorrefmark{1}, 
Huawei Cao\IEEEauthorrefmark{1}\IEEEauthorrefmark{2}\IEEEauthorrefmark{3}\thanks{\IEEEauthorrefmark{3} Corresponding author},
Ning Lin\IEEEauthorrefmark{4},
Xiaochun Ye\IEEEauthorrefmark{1} and Dongrui Fan\IEEEauthorrefmark{1}}
\IEEEauthorblockA{\IEEEauthorrefmark{1}Institute of Computing Technology, Chinese Academy of Sciences, China\\
\IEEEauthorrefmark{2}Zhongguancun Laboratory, China\\
\IEEEauthorrefmark{4}The University of Hong Kong, China\\
Email: \{xuanwei, liangyan, caohuawei, yexiaochun, fandr\}@ict.ac.cn, linning@hku.hk}}

\maketitle

\begin{abstract}

Triangle counting is a fundamental problem in graph mining, essential for analyzing graph streams with arbitrary edge orders. However, exact counting becomes impractical due to the massive size of real-world graph streams. To address this, approximate algorithms have been developed, but existing distributed streaming algorithms lack adaptability and struggle with edge deletions. In this article, we propose DTC, a novel family of single-pass distributed streaming algorithms for global and local triangle counting in fully dynamic graph streams. Our DTC-AR algorithm accurately estimates triangle counts without prior knowledge of graph size, leveraging multi-machine resources. Additionally, we introduce DTC-FD, an algorithm tailored for fully dynamic graph streams, incorporating edge insertions and deletions. Using Random Pairing and future edge insertion compensation, DTC-FD achieves unbiased and accurate approximations across multiple machines. Experimental results demonstrate significant improvements over baselines. DTC-AR achieves up to $2029.4\times$ and $27.1\times$ more accuracy, while maintaining the best trade-off between accuracy and storage space. DTC-FD reduces estimation errors by up to $32.5\times$ and $19.3\times$, scaling linearly with graph stream size. These findings highlight the effectiveness of our proposed algorithms in tackling triangle counting in real-world scenarios. The source code and datasets are released and available at \href{https://github.com/wayne4s/srds-dtc.git}{https://github.com/wayne4s/srds-dtc.git}.

\end{abstract}

\section{Introduction}\label{sec-introduction}

In the realm of massive and rapid graph streams, a crucial and valuable task entails summarizing the structure of these streams to facilitate efficient and effective graph query processing \cite{jia2023persistent,christopoulos2023local}. Triangle counting, which involves identifying sets of three pairwise interconnected edges, is a fundamental problem in graph mining with wide-ranging applications in the real world. It plays a crucial role in various scenarios such as anomaly detection \cite{lim2018memory}, spam detection \cite{becchetti2008efficient}, and social network analysis \cite{leskovec2008microscopic}. Furthermore, the count of global and local triangles serves as a vital component in graph theory, contributing to metrics such as transitive ratio \cite{newman2003structure}, clustering coefficients \cite{assadi2023streaming}, and connectivity \cite{foucault2010friend}. Graphs, in real-world contexts, often consist of a continuous stream of edges known as graph streams. These streams require real-time processing to keep up with the rapidly changing edge data. As such, efficient algorithms for triangle counting are essential to handle the dynamic nature of graph streams and extract meaningful insights.

\begin{figure}
    \centering
    \includegraphics[width=.98\linewidth]{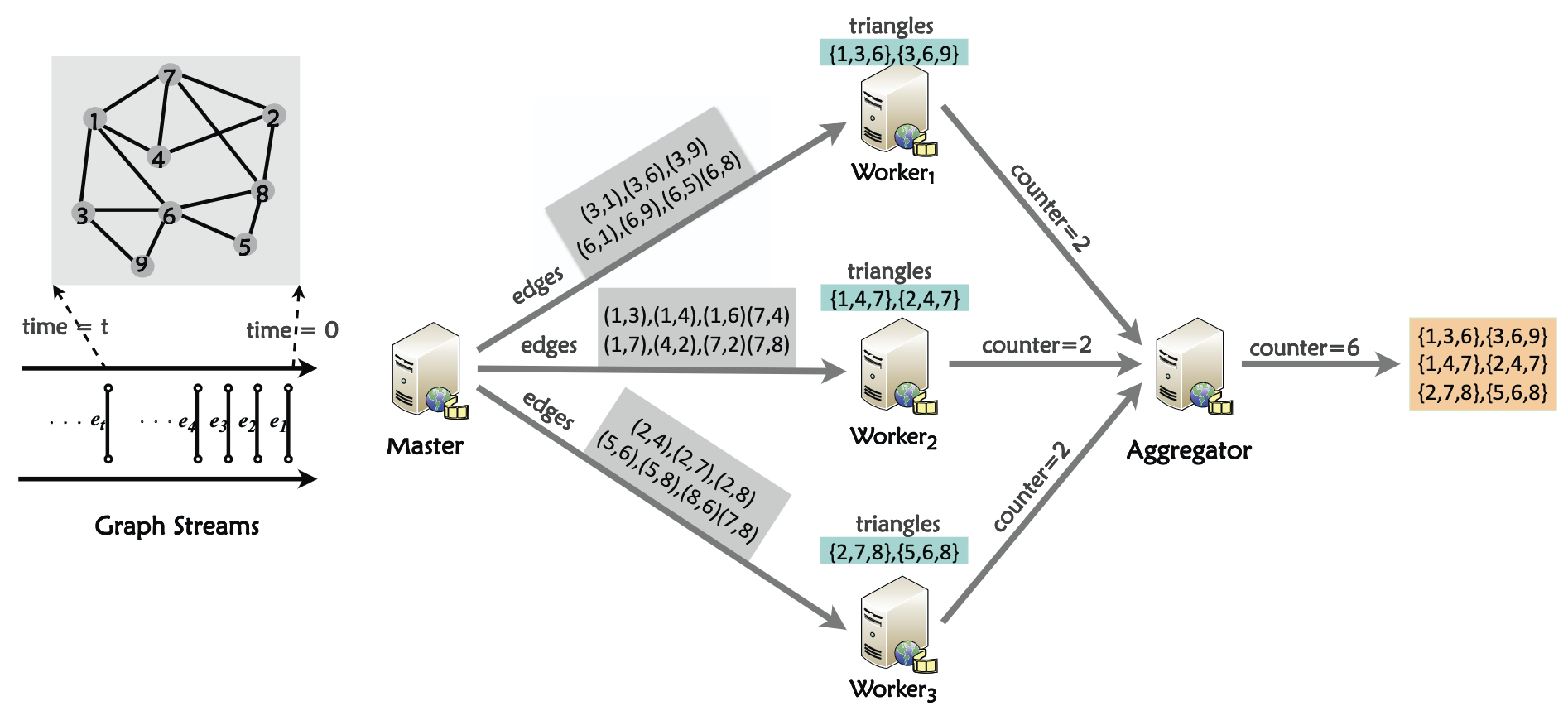}
    \caption{An illustrative architecture of distributed triangle counting in graph streams.}
    \label{figs:intro_dist_arch}
\end{figure}

Efficient implementation of triangle counting algorithms in real-world graph streams poses several challenges. \textit{Firstly, the scale of graph streams is immense, making it difficult to handle them effectively. Secondly, the lack of prior knowledge about graph streams, including their precise size, presents a challenge as they continuously evolve over time. Lastly, the limited memory space of computers restricts the ability to store the entire graph streams.} Given these challenges, obtaining exact counts of global and local triangles by storing the entire graph streams in memory becomes practically impossible. Consequently, recent research has focused on streaming algorithms that employ sampling approaches for obtaining approximate triangle counts \cite{pavan2013counting,ahmed2014graph,stefani2017triest,ahmed2017sampling,lim2018memory,pagh2012colorful}. These algorithms maintain a memory budget for uniformly sampled edges and gradually update triangle counts based on these samples, reflecting the dynamic nature of evolving graph streams. By dynamically maintaining and updating sampled edges within the memory budget, streaming algorithms avoid the impracticality of storing the entire graph streams. However, existing streaming algorithms have limitations and require further improvements to address the challenges effectively. In this paper, we propose novel approaches to overcome these limitations and enhance triangle counting in real-world graph streams. Our methods aim to achieve more accurate and efficient estimation of triangle counts while efficiently utilizing available memory resources.

\begin{table*}
    \centering
    \caption{Comparison of triangle counting algorithms in graph streams. Our proposed DTC-family outperforms state-of-the-art approaches in terms of supporting a broader range of criteria}
    \label{tab:comparison}
    \begin{tabular}{lccccccc}
        \toprule
        Criteria &DOULION,CTS &\makecell[c]{WRS \\ TRIEST-IMPR}&\makecell[c]{TRIEST-FD \\ WRS$_{DEL}$,ThinkD}&\makecell[c]{BSR-TC}&\makecell[c]{Tri-Fly,DVHT \\ DEHT,CoCoS}&{DTC-family}\\
        \midrule
        Support Edge Insertions & {\footnotesize \cmark} & {\footnotesize \cmark} & {\footnotesize \cmark} & {\footnotesize \cmark} & {\footnotesize \cmark} & {\bf \large \cmark} \\
        Controllable Memory Budget & {\ding{55}} & {\footnotesize \cmark} & {\footnotesize \cmark} & {\footnotesize \cmark} &{\footnotesize \cmark}  &  {\bf \large \cmark} \\
        No Secondary Storage & {\footnotesize \cmark} & {\footnotesize \cmark} & {\footnotesize \cmark} & {\footnotesize \cmark} & {\footnotesize \cmark}  &  {\bf \large \cmark} \\
        Single Pass Processing & {\ding{55}} & {\footnotesize \cmark} & {\footnotesize \cmark} & {\footnotesize \cmark} & {\footnotesize \cmark}  &  {\bf \large \cmark} \\
        Count Global \& Local Triangles & {\ding{55}} & {\footnotesize \cmark} & {\footnotesize \cmark} & {\footnotesize \cmark} & {\footnotesize \cmark} &  {\bf \large \cmark} \\
        Utilize Multiple Machines & {\ding{55}} & {\ding{55}} & {\ding{55}} & {\ding{55}} & {\footnotesize \cmark} &  {\bf \large \cmark} \\
        Adaptive Sampling Ratio & {\ding{55}} & {\ding{55}} & {\ding{55}} & {\footnotesize \cmark} & {\ding{55}} & {\bf \large \cmark}   \\
        Support Edge Deletions & {\ding{55}} & {\ding{55}} & {\footnotesize \cmark} & {\ding{55}} & {\ding{55}}  & {\bf \large \cmark} \\
        \bottomrule
    \end{tabular}
\end{table*}

Traditionally, streaming algorithms have primarily focused on single-machine implementations, such as Reservoir-based sampling methods \cite{vitter1985random,gemulla2008maintaining,pavan2013counting,kavassery2018improved,shin2017wrs,shin2020fast} and Bernoulli-based sampling methods \cite{tsourakakis2009doulion,pagh2012colorful,mcgregor2016better,ahmed2014graph,ahmed2017sampling}, as categorized by Xuan et al. \cite{xuantriangle}. \textit{These algorithms operate solely on a single machine by employing various sampling techniques to approximate triangle counts.} To fully harness the computational power and memory resources available across multiple machines, researchers have explored distributed streaming algorithms for triangle counting, as depicted in Figure~\ref{figs:intro_dist_arch}. Examples include Try-fly \cite{shin2018tri}, DVHT \& DEHT \cite{yu2019distributed,yang2022distributed}, and CoCoS \cite{shin2021cocos}. \textit{However, these algorithms are limited to handling edge insertions and require prior knowledge of the entire graph stream size to set appropriate parameters for accurate estimation.}

In this article, we present DTC-family ($\mathbf{D}$istributed $\mathbf{T}$riangle $\mathbf{C}$ounting), a suite of streaming algorithms designed for counting global and local triangles in fully dynamic graph streams, encompassing both edge insertions and deletions. By introducing the DTC-family, our research provides novel solutions for efficient triangle counting in dynamic graph streams, addressing the challenges of accuracy and scalability. Our proposed distributed streaming algorithms offer the following contributions:

\begin{itemize}
    \item Proposed DTC-AR, an adaptive sampling ratio method that achieves high accuracy in distributed triangle counting, eliminating the need for prior knowledge about the graph streams.
    \item Proposed DTC-FD estimates the number of triangles in fully dynamic graph streams, considering edge insertions and deletions across multiple machines.
    \item Through extensive experimentation on real-world datasets, our results demonstrate the effectiveness of our algorithms: (1) DTC-AR enables accurate and stable triangle counts in multiple machines, using a specified sampling ratio threshold. (2) DTC-FD handles triangle counting in fully dynamic graph streams using a distributed streaming schema.
\end{itemize}

The article is structured as follows. Section~\ref{sec-related-work} provides an overview of existing research on triangle counting in graph streams, covering both single-machine and multi-machine algorithms. In Section~\ref{sec-prelimi}, we present the preliminaries and motivations that underpin our proposed methods. The detailed design approaches are elucidated in Section~\ref{sec-algo}. We then conduct extensive experimentation on real-world datasets in Section~\ref{sec-expe}. Finally, Section~\ref{sec-con} concludes the article, summarizing the key findings and contributions of our research.
\section{Related Work}\label{sec-related-work}

Significant research efforts have been dedicated to triangle counting, encompassing a wide range of approaches such as exact algorithms~\cite{arifuzzaman2015space,bisson2017high}, MapReduce algorithms~\cite{pagh2012colorful,park2013efficient}, GPU-based algorithms~\cite{hu2018tricore,hu2021accelerating}, and multi-core algorithms~\cite{rahman2013approximate,shun2015multicore}. In this section, we provide a comprehensive review of the existing literature pertaining to triangle counting in graph streams.
Specifically, we highlight the findings of Table~\ref{tab:comparison}, which presents a comparative analysis of different streaming algorithms utilized for triangle counting in evolving graph streams.
This analysis provides valuable insights into the strengths and limitations of these algorithms, offering a deeper understanding of the gaps and challenges that motivate our own research.

\subsection{Single-Machine Streaming Algorithms}

In this section, we conduct a thorough and comprehensive review of the current research on single-machine triangle counting in graph streams. Our analysis is centered around four key aspects, allowing us to delve into the intricacies of the existing approaches. To facilitate our analysis, we introduce two fundamental categories: Reservoir-Based and Bernoulli-Based Sampling Patterns, as previously defined in~\cite{xuantriangle}. These categories serve as a framework for understanding and evaluating the existing approaches in triangle counting. By examining and analyzing the research contributions in these areas, we aim to gain valuable insights into the advancements made and the challenges that remain in the field of triangle counting in graph streams.

\subsubsection{Reservoir-Based Sampling}

This section focuses on the sampling method that employs a fixed-size storage budget to store sampled edges for triangle counting. Considering the limited memory space available in computers, Efficient utilization of available storage resources is crucial in this context. The standard reservoir sampling technique was first introduced by Vitter~\cite{vitter1985random}. Gemulla et al. further proposed a novel sampling method called Random Pairing (RP), which handled both edge insertions and deletions for triangle counting in graph streams~\cite{gemulla2008maintaining}. Pavan et al. presented a neighborhood sampling algorithm, leveraging standard reservoir sampling for global triangle counting in graph streams~\cite{pavan2013counting}. Based on the neighborhood sampling algorithm above, Kavassery et al. further proposed an extended algorithm, known as Neighborhood Multisampling (NMS), to obtain higher accuracy for triangle counting by a multi-sampling approach\cite{kavassery2018improved}. Additionally, Shin et al. proposed a single-pass streaming algorithm called Waiting-Room Sampling (WRS)~\cite{shin2017wrs}. By exploiting the temporal locality of real-world graphs and utilizing standard reservoir sampling, WRS achieves efficient triangle counting in a streaming setting.

\subsubsection{Bernoulli-Based Sampling}

Instead of setting storage budget with a fixed size by Reservoir-Based Sampling, Bernoulli-Based Sampling begins with a given probability $p$, and monotonically increases storage space consumed by sampled edges in involving graph streams. Tsourakakis et al. \cite{tsourakakis2009doulion} proposed an algorithm called DOULION, sampling each edge independently with same probability $p$ and updating the count with $1/p^3$. Another approach is the colorful triangle sampling (CTS) algorithm proposed by Pagh and Tsourakakis~\cite{ pagh2012colorful}. This algorithm assigns a random color to each vertex and samples only the edges that have the same color on both endpoints. The number of triangles is then estimated based on the sampled monochromatic edges. McGregor et al. introduced space-efficient data stream algorithms for counting triangles in two different models~\cite{mcgregor2016better}. Based on a Horvitz-Thompson construction \cite{horvitz1952generalization}, Ahmed et al. proposed a generic sampling framework known as Graph Sample and Hold (gSH), which is applicable to massive graph analytics in a single pass~\cite{ahmed2014graph}. gSH utilizes different sampling probabilities based on the graph properties of interest, for instance, gSH samples arriving edges without adjacencies to previously sampled edges with probability $p$ and holds other edges with probability $q$, improving the accuracy of subgraph patterns estimation. Lim et al.~\cite{lim2015mascot} presented a memory-efficient and accurate one-pass method named Memory-efficient Accurate Sampling for Counting Local Triangles (MASCOT). MASCOT adopts a strategy of $unconditional$ $counting$ $before$ $sampling$, providing an effective approach for local triangle estimation over graph streams.

\subsubsection{Handle Insertion-Only Edge Streams}

Tsourakakis et al.~\cite{tsourakakis2008fast} and Kolountzakis et al.~\cite{kolountzakis2012efficient} used multi-pass algorithms to count the number of global triangles, respectively. Ahmed et al. introduced a sample and hold framework called gSH \cite{ahmed2014graph}, which is specifically designed for estimating subgraphs of interest, such as triangles, in graph streams that only have edge insertions. Further, Jha et al. proposed a single-pass method for estimating the count of global triangles using the classic probabilistic result of the birthday paradox\cite{jha2015space}. Based on reservoir sampling, De Stefani et al. \cite{stefani2017triest} proposed TRIEST-IMPR, which improves estimate accuray by fully utilizing storage budget in insertion-only graph streams. 

\subsubsection{Handle Fully Dynamic Streams}

Kutzkov and Pagh~\cite{kutzkov2014triangle} proposed the first single-pass algorithm for triangle counting in fully dynamic graph streams. However, this algorithm is not suitable for handling real-world graph streams because it computes the estimation only at the end of the stream, which can be computationally expensive. To address this issue, De Stefani et al. proposed TRIEST-FD~\cite{stefani2017triest}, a method for estimating the number of triangles in fully dynamic graph streams. TRIEST-FD utilizes Random Pairing (RP)~\cite{gemulla2008maintaining} and takes advantage of future edge insertions to compensate for edge deletions, improving the accuracy of triangle count estimation. Instead of simply discarding unsampled edges, Shin et al.~\cite{shin2018think} proposed alternative algorithms called Think before you Discard (ThinkD). ThinkD leverages those unsampled edges to update triangle counts before discarding them. By exploring the unsampled edges, ThinkD achieves better accurate estimation of both global and local triangles compared to TRIEST-FD, while utilizing the same storage budget.

\subsection{Multi-Machine Distributed Streaming Algorithms}

To harness the computational power and storage resources of multiple machines, researchers have explored distributed streaming algorithms for approximate triangle counting in evolving graph streams. Pavan et al.~\cite{pavan2013parallel} were among the first to propose distributed streaming algorithms for estimating the number of triangles across multiple machines, which laid the foundation for distributed triangle counting algorithms. Wang et al.~\cite{wang2019rept} introduced a parallel single-pass streaming algorithm called REPT to reduce the covariance between sampled triangles. Building upon TRIEST-BASE and TRIEST-IMPR~\cite{stefani2017triest}, Yu et al.~\cite{yu2019distributed} proposed two approximate counting algorithms for the global triangles in a distributed environment, and later extended and refined them in~\cite{yang2022distributed}. Shin et al. presented Tri-Fly~\cite{shin2018tri}, a distributed streaming algorithm for both global and local triangle counting. Tri-Fly utilizes a parallel implementation of TRIEST-IMPR~\cite{stefani2017triest} and employs a broadcast scheduling strategy to send all edges to every worker, which ensures that each worker has access to the complete set of edges for accurate triangle counting. Based on Tri-Fly, Shin et al.~\cite{shin2021cocos} further introduced optimized CoCoS, which reduces redundancy in storage and guarantees the absence of disintegrated triangles, contributing to more efficient triangle counting in distributed graph streams.
\section{Preliminaries and Motivations}\label{sec-prelimi}

This section aims to establish a solid understanding of the core concepts and motivations, setting the stage for the subsequent discussions and providing a comprehensive context for the research presented in this article.

\subsection{Definitions}

Table~\ref{tab:definitions} provides a list of symbols and definitions that are frequently used throughout this article. Consider an undirected graph stream $\{e_{0}, e_{1}, e_{2},...\}$, where each $e_i$ represents an edge in the stream. We define $\mathcal{G}_t$ = ($\mathcal{V}_t$, $\mathcal{E}_t$) as the graph composed of edges up to time $t$. Here, $\mathcal{V}_t$ represents the set of vertices, and $\mathcal{E}_t$ represents the set of edges that have arrived by time $t$. A triangle $\{u,v,c\}$ refers to an unordered triple consisting of nodes $u$, $v$, and $c$, where each pair is connected by an edge. In other words, there exist edges between $u$ and $v$, $v$ and $c$, and $c$ and $u$. For fully dynamic graph streams, we denote an arriving edge as $\{u,v,\odot\}$, where $\odot \in \{+,-\}$ represents whether the edge is being inserted or deleted, respectively. Here, $u$ and $v$ represent two distinct vertices connected by the edge. These definitions lay the groundwork for discussing the subsequent topics related to graph stream analysis and provide a clear understanding of the terminology used throughout the article.

\begin{table}[t]
    \centering
    \small
    \caption{A comprehensive list of frequently involved symbols \label{tab:definitions}}
    \begin{tabular}{c|l}
        \toprule
        \textbf{Symbol} & \textbf{Definition} \\
        
        \midrule
        \multicolumn{2}{l}{Definitions for Graph Streams} \\ 
        \midrule
        {  $\mathcal{G}$$_t$ = ($\mathcal{V}$$_t$,$\mathcal{E}$$_t$) } & {graph $\mathcal{G}$$_t$ at time $t$ }\\
        $\{e_{0}, e_{1}, e_{2},...\}$ & input graph stream \\
        $e_{t}$ & the $t$-th arriving edge ($t\in\{0,1,2,...\}$)\\
    
        $\{u,v,\odot\}$ & fully dynamic edge ($\odot\in\{+,-\}$) \\
        $\{u,v,c\}$ & triangle formed by nodes $u$, $v$ and $c$ \\
        
        \midrule
        \multicolumn{2}{l}{Definitions for Algorithms} \\
        \midrule
        $k$ & size of storage budget for each worker\\
        $\mathcal{R}$ & sampling ratio threshold for each worker \\
        $\mathcal{W}$ & number of workers \\
        $\omega_i$ & current worker \\
        $hash(u) $ & mapping function for node $u$ \\
        $\Sigma$ & maximum of available memory space \\
        $\Delta$ & estimate of global triangles by worker \\
        $\Delta_u$ & estimate of local triangles $u$ by worker \\
        $\mathcal{A}$ & estimate of global triangles by aggregator\\
        $\mathcal{A}_u$ & estimate of local triangles $u$ by aggregator\\
        $\mathcal{S}$ & set of sampled edges \\
        $\mathcal{N}^{\mathcal{S}}_{u}$ &  set of neighbors of node $u$ in $\mathcal{S}$ \\
        \bottomrule
    \end{tabular}
\end{table}

\subsection{Motivations}

Triangle counting in graph streams is a fundamental problem with numerous applications in network analysis, social network analysis, and recommendation systems. Existing distributed streaming algorithms for triangle counting face several challenges that hinder their scalability and accuracy. In this paper, we aim to address these challenges and present a comprehensive solution for accurate and efficient triangle counting in fully dynamic graph streams.

1) $\textit{Requirement of prior knowledge}$: Current algorithms heavily rely on prior knowledge to determine appropriate parameters, such as the initial storage budget, for achieving an optimal balance between accuracy and speed. However, obtaining such prior knowledge about evolving graph streams is often impractical or even impossible. Consequently, selecting appropriate parameters to achieve an optimal trade-off between accuracy and speed becomes a significant challenge.

2) $\textit{Limitation to insertion-only graph streams}$: Many existing distributed streaming algorithms for triangle counting are designed to handle graph streams that only involve edge insertions. They overlook the fact that real-world graph streams are fully dynamic and may include both edge insertions and deletions. Ignoring the impact of edge deletions in the streaming process limits the algorithms' effectiveness and hinders their ability to capture the evolving graph's structural changes accurately.

3) $\textit{Lack of efficient distributed streaming algorithms}$: Most research efforts in triangle counting have primarily focused on single-machine algorithms, neglecting the distributed nature of the problem. While single-machine algorithms provide valuable insights, they are not scalable for large graph streams or distributed environments. The lack of efficient distributed streaming algorithms for triangle counting further exacerbates the challenge of processing graph streams in a distributed setting. Additionally, there is a lack of efficient distributed streaming algorithms that can handle triangle counting without prior knowledge or in fully dynamic streaming graph settings.

To overcome these challenges, we propose the DTC-family, a suite of distributed streaming algorithms specifically tailored for global and local triangle counting in fully dynamic graph streams. The DTC-family algorithms aim to address the limitations of prior approaches and provide efficient solutions for accurate and scalable triangle counting in dynamic distributed environments. By developing the DTC-family algorithms, we aim to enable efficient triangle counting in fully dynamic graph streams without requiring prior knowledge and accommodate both edge insertions and deletions. Our algorithms leverage novel sampling techniques, such as Reservoir-based and Bernoulli-based sampling, to strike a balance between accuracy and memory consumption. We also introduce distributed strategies that effectively handle the challenges posed by the distributed nature of graph stream processing.

Through extensive experimentation and comparative analysis, we demonstrate the superior performance of the DTC-family algorithms in terms of estimation accuracy, memory usage, and computational efficiency. The results highlight the effectiveness of our proposed algorithms and their potential for practical applications in various domains that rely on triangle counting in dynamic distributed graph streams.
\section{Proposed Algorithms}\label{sec-algo}

In this section, we present an overview of the classical architecture of distributed streaming algorithms for approximative triangle counting, known as Master-Workers-Aggregator (MWA). Subsequently, we propose the DTC ($\mathbf{D}$istributed $\mathbf{T}$riangle $\mathbf{C}$ounting), a family of distributed streaming algorithms to estimate global and local triangles in graph streams. Specifically, we introduce two variants of the DTC algorithm: DTC-AR for accurate estimation of triangles in insertion-only edges without prior knowledge, and DTC-FD, which employs Random Pairing (RP)~\cite{gemulla2008maintaining} to handle triangle counting in fully-dynamic graph streams involving both edge insertions and deletions.

\subsection{Architecture of MWA} \label{architecture:wma}

The classical architecture for approximative triangle counting in distributed streaming algorithms is known as Master-Workers-Aggregator (MWA). This architecture involves a master node responsible for coordinating the computation and multiple worker nodes that process data in parallel. The aggregator node consolidates the results from the workers and provides the final estimation of triangle counts. The master node distributes edges among the workers using a schedule strategy, which involves comparing the hash values of vertices $u$ and $v$ in an edge $\{u,v\}$. If $hash(u)$ matches $hash(v)$, the master unicasts the edge $\{u,v\}$ to the worker with the corresponding hash value. Alternatively, if $hash(u)$ and $hash(v)$ differ, the master broadcasts the edge $\{u,v\}$ to all workers. Each worker, denoted as $\omega_i$, unconditionally updates the triangle count and selectively stores the edge $\{u,v\}$ with a non-zero probability when the hash value of $u$ (or $v$) matches $\omega_i$. Finally, the aggregator node accumulates triangle counts from all workers in real time. This architecture typically employs multiple workers, while maintaining a single master and aggregator.

Figure~\ref{figs:intro_dist_arch} illustrates the proposed MWA architecture for estimating the number of global triangles. The configuration consists of one master node, three workers (with enough storage budget), and one aggregator. In this specific scenario, the master assigns each edge $\{u,v\}$ based on the hash values of $u$ and $v$. Each worker individually counts two triangles. The aggregator node receives and accumulates updates from all workers to derive the final triangle count estimation.

\subsection {DTC-AR Algorithm}  

We first introduces DTC-AR, a novel distributed streaming algorithm designed for accurate global and local triangle counting in evolving graph streams. DTC-AR employs an adaptive approach to dynamically resample the storage budget upwards, ensuring efficient resource utilization. Our proposed Algorithm~\ref{alg:dtc-ar} for DTC-AR is illustrated below.

$\bullet~\mathbf{Master}$ (lines~\ref{alg:dtc-ar:master:start}-\ref{alg:dtc-ar:master:end} of Algorithm~\ref{alg:dtc-ar}): As shown in line \ref{alg:dtc-ar:master:schedule} of Algorithm \ref{alg:dtc-ar}, we use the function \normalsize{S}\scriptsize{CHEDULE}\normalsize{E}\scriptsize{\textsc{DGE}}  \normalsize{($\{u,v,\odot\}$)} to distribute every edge from the master to workers. After receiving an edge $e_t\{u,v,+\}$ from evolving graph streams, the master will first 
propagate it to one or more workers by a mapping function $hash$ of two nodes $u$ and $v$. If $hash(u)$ is equal to $hash(v)$ (line~\ref{alg:dtc-ar:master:unicast-start}), the nodes $u$ and $v$ are mapped to same worker. Then, the master only unicasts this edge $e_t\{u,v,+\}$ to the worker (indexed by $hash(u)$ or $hash(v)$) (line \ref{alg:dtc-ar:master:unicast}). Otherwise, the master will broadcast this edge $e_t\{u,v,+\}$ to all workers (line \ref{alg:dtc-ar:master:broadcast}).

$\bullet~\mathbf{Worker}$ (lines~\ref{alg:dtc-ar:worker:start}-\ref{alg:dtc-ar:worker:end} of Algorithm~\ref{alg:dtc-ar}): Each worker independently estimates the number of global and local triangles without communicating with others. All workers begin with an empty storage budget $\mathcal{S}_c$ with size $k$ and $\mathcal{S}_p$ to store more sampled edges by adaptive sampling (line~\ref{alg:dtc-ar:worker:start}). When a worker receives an edge $\{u,v,\odot\}$ from the master, it first unconditionally updates the number of both global and local triangles (lines~\ref{alg:dtc-ar:worker:update:start}-\ref{alg:dtc-ar:worker:update:end}). For each node $c$ in common neighbor set $\mathcal{N}^{\mathcal{S}_c + \mathcal{S}_p}_{u,v}$, we update the counts by $1/p_{ar}$. However, DTC-AR only samples the edge $\{u,v\}$ depending on the values of $hash(u)$ and $hash(v)$ (lines~\ref{alg:dtc-ar:worker:sample:start}-\ref{alg:dtc-ar:worker:sample:end}). Here, only when $hash(u)$ (or $hash(v)$) is equal to $\omega_i$, DTC-AR calls $\textsc{SampleInsEdge}$ to sample the edge $\{u,v,+\}$ with non-zero probability (line~\ref{alg:dtc-ar:worker:sample:reservoir}). As shown in lines \ref{alg:dtc-ar:worker:sampleIns:start}-\ref{alg:dtc-ar:worker:sampleIns:end} of Algorithm~\ref{alg:dtc-ar}, we implement the sampling algorithm $\textsc{SampleInsEdge}$ based on standard reservoir sampling \cite{vitter1985random}.

\begin{algorithm}[t]
    \DontPrintSemicolon
    \SetKwInOut{Input}{\textbf{Input}}	
    \SetKwInOut{Output}{\textbf{Output}}
    
    \SetKwFunction{algo}{algo}
    \SetKwFunction{proc}{\textbf{Procedure}}{}{}
    \SetKwFunction{Master}{\textbf{Master}}{}{}
    \SetKwFunction{Worker}{\textbf{Worker}}{}{}
    \SetKwFunction{Aggregator}{\textbf{Aggregator}}{}{}
    \SetKwFunction{main}{\textbf{ChooseCenter}}
    \SetKwFunction{quant}{\textbf{CalculateUniqueness}}
    \Input{(1) \{e$_{0}$, e$_{1}$,...\}: insertion-only graph streams;\\
        (2) $k$: storage budget for each worker;\\
        (3) $\mathcal{R}$: sampling ratio threshold.
    }
    \Output{$\mathcal{A}$, $\mathcal{A}_u$\\
    }

    \vspace{1mm}
    \Master:\\ 
    
    \For{each arriving edge $e_t\{u,v,\odot\}$} {\label{alg:dtc-ar:master:start}
        \normalsize{S}\scriptsize{CHEDULE}\normalsize{E}\scriptsize{\textsc{DGE}}  \normalsize{($\{u,v,\odot\}$)} \label{alg:dtc-ar:master:schedule}
    }
    \SetKwBlock{Function}{\texttt{function}   \textsc{ScheduleEdge}($\{u,v,\odot\}$) }{end}	
    \Function{     
            \If{hash(u) = hash(v)}{ \label{alg:dtc-ar:master:unicast-start}
                unicast $e_t\{u,v,\odot\}$ to worker $hash(u)$\label{alg:dtc-ar:master:unicast}
            }\label{alg:dtc-ar:master:unicast-end}
            \Else { \label{alg:dtc-ar:master:broadcast-start}
                broadcast $e_t\{u,v,\odot\}$ to every worker \label{alg:dtc-ar:master:broadcast}
            } \label{alg:dtc-ar:master:broadcast-end}    
        } \label{alg:dtc-ar:master:end}
	
    \vspace{1mm}
    \Worker:\\ 
        $\mathcal{S}_c \leftarrowtail \emptyset , ~\mathcal{S}_p \leftarrowtail \emptyset,~ \Delta \leftarrowtail 0, ~\Delta_u \leftarrowtail 0,~t \leftarrowtail 0$\\ \label{alg:dtc-ar:worker:start}
        
        \For{each arriving edge $e_t\{u,v,\odot\}$ from Master} { 
            $t \leftarrowtail t \odot 1$\\
            \ForEach{ $  c \in \mathcal{N}^{\mathcal{S}_c + \mathcal{S}_p}_{u} \cap \mathcal{N}^{\mathcal{S}_c + \mathcal{S}_p}_{v}  $}{  \label{alg:dtc-ar:worker:update:start} 
                $\Delta_u \leftarrowtail \Delta_u \odot 1/p_{ar},$
                $\Delta_v \leftarrowtail \Delta_v \odot 1/p_{ar}$\\
                $\Delta_c \leftarrowtail \Delta_c \odot 1/p_{ar},$
                $\Delta \leftarrowtail \Delta \odot 1/p_{ar}$\\
            }\label{alg:dtc-ar:worker:update:end}
            
            \If {hash(u) = $\omega_i$ or hash(v) = $\omega_i$} {  \label{alg:dtc-ar:worker:sample:start} 
                \textsc{SampleInsEdge}($\mathcal{S}_c, \{u,v,\odot\}$)\\ \label{alg:dtc-ar:worker:sample:reservoir}
                \textsc{CheckRatio}() \label{alg:dtc-ar:worker:sample:check} 
            }\label{alg:dtc-ar:worker:sample:end}	
        }
		
    \SetKwBlock{Function}{\texttt{function} \textsc{SampleInsEdge}($\mathcal{S}, \{u,v,\odot\}$) }{end}
    \Function{ \label{alg:dtc-ar:worker:sampleIns:start} 
        \If{$\lvert \mathcal{S} \rvert < k$}{
            $\mathcal{S} \leftarrowtail \mathcal{S} \cup \{u,v\}$
        }
        \ElseIf{$k/t < a~random~number(0,1)$} {
            choose a random edge $\{m,n\}$ from $\mathcal{S}$ \\
            $\mathcal{S} \leftarrowtail \mathcal{S} \setminus \{m,n\}$ \\
            $\mathcal{S} \leftarrowtail \mathcal{S} \cup \{u,v\}$ \\
        }	
    } \label{alg:dtc-ar:worker:sampleIns:end}    
    
    \SetKwBlock{Function}{\texttt{function}
        \textsc {CheckRatio}()
    }{end}
    \Function{\label{alg:dtc-ar:check:start}
        
        \If {$r_i = \mathcal{R}$ and $\Sigma \geq k$ } { 
            
            $\mathcal{S}_p \leftarrowtail \mathcal{S}_p \cup \mathcal{S}_c$;\\ 
            create a new $\mathcal{S}_c$ with size $k$\\  
            $\Sigma \leftarrow \Sigma - k, ~t \leftarrowtail 0$
        }		
    }\label{alg:dtc-ar:check:end} \label{alg:dtc-ar:worker:end}

    \vspace{1mm}
    \Aggregator: \\
    $\mathcal{A} \leftarrowtail 0$,  $\mathcal{A}_u \leftarrowtail 0$\\ \label{alg:dtc-ar:aggregator:start}
    \For{each $\Delta$ and $\Delta_u$ from workers} {
        \textsc{AccumulateCount}() \label{alg:dtc-ar:aggregator:count:call}
    }
    
    \SetKwBlock{Function}{\texttt{function} \textsc{AccumulateCount}() }{end}
    \Function{ \label{alg:dtc-ar:aggregator:count:start}
            $\mathcal{A} \leftarrowtail \mathcal{A} + \Delta$\\
            $\mathcal{A}_u \leftarrowtail \mathcal{A}_u + \Delta_u$
        
    }\label{alg:dtc-ar:aggregator:end} \label{alg:dtc-ar:aggregator:count:end} 
    \caption{DTC-AR}\label{alg:dtc-ar}
\end{algorithm}

Then, \textsc{CheckRatio} is called to self-adaptively update storage budget to maintain the current sampling ratio greater than or equal to the given threshold $\mathcal{R}$ (lines~\ref{alg:dtc-ar:check:start}-\ref{alg:dtc-ar:check:end}). In function \textsc{CheckRatio}, $\Sigma$ is the totally available space for every worker. $\Sigma$ is further subdivided into $\mathcal{S}_c$, which represents the current sampling set with probability $r_i$, and $\mathcal{S}_p$ is a sampling pool of constant probability $\mathcal{R}$. For the convenience of description, we presume that every allocated sampling set $\mathcal{S}_c$ is initialized with sample size $k$ in all workers. Once the sampling ratio of worker $\omega_i$ is equivalent to $\mathcal{R}$ specified by users, and $\Sigma$ is no less than size $k$, $\mathcal{S}_c$ will be integrated into $\mathcal{S}_p$. Then, a new storage budget $\mathcal{S}_c$ with same size $k$ is allocated, and the available space of $\Sigma$ decreases. Subsequently, DTC-AR restarts a new round standard reservoir sampling from scratch again. Based on this method, DTC-AR enables the sampling ratio of every worker greater than or equal to a given $\mathcal{R}$ in distributed computer clusters, without any knowledge about the size of evolving graph streams.

$\bullet~\mathbf{Aggregator}$ (lines~\ref{alg:dtc-ar:aggregator:start}-\ref{alg:dtc-ar:aggregator:end} of Algorithm~\ref{alg:dtc-ar}): Whenever it receives the updates of global and local triangles from workers, the aggregator accumulates them to obtain up-to-date counts by calling the function \textsc{AccumulateCount} (line \ref{alg:dtc-ar:aggregator:count:call}). Every triangle can be counted with non-zero probability in one worker at most by the unicast and broadcast strategy In the $\mathbf{MWA}$ (Section~\ref{architecture:wma}). The aggregator can obtain global triangle count $\mathcal{A}$ and local triangle count $\mathcal{A}_u$ for node $u$ at any time $t$.

$\bullet~\mathbf{Theoretical}~\mathbf{Analysis}$

Whenever a new arriving edge $\{u,v\}$ forms a triangle $\{u,v,c\}$  with edges $\{u,c\}$ and $\{v,c\}$ in $\mathcal{S}_c \cup \mathcal{S}_p$, we need to calculate the sampling probability. Depending on the positions ($\mathcal{S}_c$ or $\mathcal{S}_p$) of edge $\{u,c\}$ and $\{v,c\}$, the probability $p_{ar}$ is divided into four types as follows.

{\small 
    \begin{equation} \label{equal:type}
    p_{ar} = \begin{cases} 
    min\{1, \frac{k}{t} \times \frac{k-1}{t-1}\} & \textnormal{type-1}\  \\
    \mathcal{R} \times \frac{k-1}{(k/\mathcal{R})-1} & \textnormal{type-2}\ \\
    \mathcal{R} \times \mathcal{R} & \textnormal{type-3}\ \\
    min\{1, \frac{k}{t}\} \times \mathcal{R} & \textnormal{type-4}\ \\
    \end{cases}
    \end{equation}
}

\textbf{Type-1} ($\{$\{u,c\}$, $\{v,c\}$ \} \subset \mathcal{S}_c$): When $k < \lvert \mathcal{S}_c \rvert $, the probabilities of sampling  $(u, c)$ and $(v, c)$ stored in $\mathcal{S}_c$ are $k/{t}$ and $(k-1)/({t-1})$, respectively. Otherwise, when $\lvert \mathcal{S}_c \rvert \leq k$, the probabilities are equal to $1$.

\textbf{Type-2} ($\{ $\{u,c\}$, $\{v,c\}$ \} \subset \mathcal{S}^{(i)}_{p}$): When both edges $\{u,c\}$ and $\{v,c\}$ are located in same $\mathcal{S}^{(i)}_{p}$ of the sampling pool $i$, the probabilities of sampling them are ${\mathcal{R}}$ and ${(k-1)}/{(k/\mathcal{R}-1)}$, respectively.

\textbf{Type-3} ($\{ $\{u,c\}$, $\{v,c\}$ \} \subset \mathcal{S}^{(i)}_{p} \cup \mathcal{S}^{(j)}_{p}, \forall i \neq j$): When the edges $\{ u, c \} \subset \mathcal{S}^{(i)}_{p}$, $\{ v, c \} \subset \mathcal{S}^{(j)}_{p}$ and $\forall i \neq j$, the probabilities of both sampled edges are ${\mathcal{R}}$.

\textbf{Type-4} ($\{$\{u,c\}$, $\{v,c\}$ \} \subset \mathcal{S}_c \cup \mathcal{S}^{(i)}_{p}$): When one edge $\{u, c\} \subset \mathcal{S}_{c}$ (or $\{v, c\} \subset \mathcal{S}_{c}$), the probability of this sampled edge is $min\{1, k/{t}\}$. And the other one $\{v, c \} \subset \mathcal{S}^{(i)}_{p}$ (or $\{u, c\} \subset \mathcal{S}^{(i)}_{p}$), the probability is $\mathcal{R}$.

Since every edge in storage budget $\mathcal{S}_c$ meets standard reservoir sampling \cite{vitter1985random} and is sampled with the same probability $\mathcal{R}$ in storage pool $\mathcal{S}_p$, DTC-AR gives an unbiased estimation of both global and local triangles. The advantage is that DTC-AR can adaptively resample to guarantee a specified bounded-sampling-ratio, without any knowledge about the size of evolving graph streams.

\begin{algorithm}[t]
    \DontPrintSemicolon
    \SetKwInOut{Input}{\textbf{Input}}	
    \SetKwInOut{Output}{\textbf{Output}}	
    \SetKwFunction{algo}{algo}
    \SetKwFunction{proc}{Procedure}{}{}
    \SetKwFunction{Master}{\textbf{Master}}{}{}
    \SetKwFunction{Worker}{\textbf{Worker}}{}{}
    \SetKwFunction{Aggregator}{\textbf{Aggregator}}{}{}
    \SetKwFunction{main}{\textbf{ChooseCenter}}
    \SetKwFunction{quant}{\textbf{CalculateUniqueness}}
    \Input{(1) \{e$_{0}$, e$_{1}$,...\}: fully dynamic graph streams;\\
        (2) $k$: memory budget for each worker.\\ 
    }
    \Output{$\mathcal{A}$, $\mathcal{A}_u$\\
    }
	
    \vspace{1mm}
    \Master: (lines~\ref{alg:dtc-ar:master:start}-\ref{alg:dtc-ar:master:end} of Algorithm~\ref{alg:dtc-ar})
    \\ 

    \vspace{1mm}
    \Worker:\\ 
    
    $\mathcal{S} \leftarrowtail \emptyset, ~\Delta \leftarrowtail 0, ~\Delta_u \leftarrowtail 0, ~t \leftarrowtail 0, ~n_g \leftarrowtail 0, n_b \leftarrowtail 0$\\ \label{alg:dtc-fd:worker:start}
    
        \For{each arriving edge e$_t$\{u,v,$\odot$\} from master} {
            $t \leftarrowtail t \odot 1$\\
            \ForEach{$ c \in \mathcal{N}^{\mathcal{S}}_u \cap \mathcal{N}^{\mathcal{S}}_v $}{ 				
                $\Delta_u \leftarrowtail \Delta_u$ $\odot$ $1/p_{fd},$
                $\Delta_v \leftarrowtail \Delta_v$ $ \odot $ $ 1/p_{fd}$\\
                $\Delta_c \leftarrowtail \Delta_c $ $\odot$ $ 1/p_{fd},$
                $\Delta \leftarrowtail \Delta$ $ \odot$ $ 1/p_{fd}$\\
            }
            
            \If {$hash(u) = \omega_i$ or $hash(v) = \omega_i$} { \label{alg:dtc-fd:worker:sample:start} 
                \textsc{SampleDynEdge}($\mathcal{S}, \{u,v,\odot\}$\\
            }\label{alg:dtc-fd:worker:sample:end}	
        }
 
    \SetKwBlock{Function}{\texttt{function} \textsc{SampleDynEdge}($\mathcal{S}, \{u,v,\odot\}$)}{end}  
    \Function{ \label{alg:dtc-fd:worker:sampleDyn:start}
        \If{$\odot = +$}{ 
            \If{$n_b + n_g = 0$ }{
                \textsc{SampleInsEdge}($\mathcal{S}, \{u,v,\odot\}$)\\
            }
            \ElseIf{$n_b / (n_g + n_b) < a~random~number(0,1)$}{
                $\mathcal{S} \leftarrowtail \mathcal{S} \cup \{u,v\}$ \\
                $n_b \leftarrowtail n_b - 1$
            }
            \Else {
            $n_g \leftarrowtail n_g - 1$
            }
        }
        \Else {
            \If{$\{u,v\} \in \mathcal{S}$}{
                $\mathcal{S} \leftarrowtail \mathcal{S} \setminus \{u,v\}, ~n_b \leftarrowtail n_b + 1$
            }
            \Else {
                $n_g \leftarrowtail n_g + 1$
            }
        } 	
    } \label{alg:dtc-fd:worker:sampleDyn:end} \label{alg:dtc-fd:worker:end}
    
    \vspace{1mm}
    \Aggregator: (lines~\ref{alg:dtc-ar:aggregator:start}-\ref{alg:dtc-ar:aggregator:end} of Algorithm~\ref{alg:dtc-ar})
        \\
    \caption{DTC-FD} \label{alg:dtc-fd}
\end{algorithm}

\subsection{DTC-FD Algorithm}

To handle fully dynamic graph streams in multiple machines for rapidly and accurately counting global and local triangles, we propose a new algorithm, namely DTC-FD, which outperforms state-of-the-art TRIEST-FD \cite{stefani2017triest} and ThinkD \cite{shin2018think} significantly. Note that $\odot \in \{+,-\}$ for fully dynamic graph streams. 

$\bullet~\mathbf{Master}$ (lines~\ref{alg:dtc-ar:master:start}-\ref{alg:dtc-ar:master:end} of Algorithm~\ref{alg:dtc-ar}): Like Algorithm~\ref{alg:dtc-ar}, the master straightforwardly calls the function \normalsize{S}\scriptsize{CHEDULE}\normalsize{E}\scriptsize{\textsc{DGE}}  \normalsize{($\{u,v,\odot\}$)} to assign each edge to one or more workers. The master distributes every edge $e_t\{u,v,\odot\}$ to workers, only by the comparison between $hash(u)$ and $hash(v)$, without regard to the sign $\odot$ of the edge $e_t$.

$\bullet~\mathbf{Worker}$ (lines~\ref{alg:dtc-fd:worker:start}-\ref{alg:dtc-fd:worker:end} of Algorithm~\ref{alg:dtc-fd}): Unlike previous algorithms \ref{alg:dtc-ar}, DTC-FD can handle fully dynamic graph streams based on the RP sampling \cite{gemulla2008maintaining}, which including both edge insertions and deletions. For each node $c$ in common neighbor set $\mathcal{N}^{\mathcal{S}}_u \cap \mathcal{N}^{\mathcal{S}}_v$, $\textsc{SampleDynEdge}$ first unconditionally updates the number of both global and local triangles (lines~\ref{alg:dtc-fd:worker:sampleDyn:start}-\ref{alg:dtc-fd:worker:sampleDyn:end}). Specifically, we increase the counts by $1/p_{fd}$ for $\odot = +$, or decrese the counts by $1/p_{fd}$ for $\odot = -$. In the case of edge insertions (i.e., $e_t\{u,v,+\}$), DTC-FD increases the number of global triangles and the corresponding local triangles of nodes $u$ and $v$, owing to new triangles formulated by edge $\{u,v\}$. Otherwise, in the case of edge deletions (i.e., $e_t\{u,v,-\}$), DTC-FD will decrease the number of global triangles and the corresponding local triangles of nodes $u$ and $v$, owing to removed triangles formulated by edge $\{u,v\}$. If and only if $hash(u)$ (or $hash(v)$) is equal to the current worker $\omega_i$, DTC-FD uses the function $\textsc{SampleDynEdge}$ to sample with probability $p_{fd}$ in fully dynamic graph streams (lines~\ref{alg:dtc-fd:worker:sample:start}-\ref{alg:dtc-fd:worker:sample:end}).

$\bullet~\mathbf{Aggregator}$ (lines~\ref{alg:dtc-ar:aggregator:start}-\ref{alg:dtc-ar:aggregator:end} of Algorithm~\ref{alg:dtc-ar}):

Like algorithms \ref{alg:dtc-ar}, the aggregator accumulates the count of global triangles $\Delta$ and the count of local triangles $\Delta_u$ from all workers to estimate the results.

$\bullet~\mathbf{Theoretical}~\mathbf{Analysis}$

Based on Random Pairing \cite{gemulla2008maintaining}, the workers update the storage budget $\mathcal{S}$ in fully dynamic streams. Generally speaking, $c_g$ and $c_b$ indicate the number of edge deletions, compensated by edge inserions in the near future. On the one hand, when a deletion of an edge $\{u,v,-\}$ arrives, RP increases $c_g$ or $c_b$ depending on the edge $\{u,v\}$ is in $\mathcal{S}$ or not. On the other hand, RP decreases $c_g$ or $c_b$ by the probability $c_b/(c_g+c_b)$ or $c_g/(c_g+c_b)$ for the edge $\{u,v,+\}$, respectively. Whenever a new arriving edge $\{u,v\}$ forms a triangle $\{u,v,c\}$  with edges $\{u,c\}$ and $\{v,c\}$ in $\mathcal{S}$, we need to calculate the sampling probability. Whatever the $\odot$ is + or -, the probability $p_{fd}$ is same as follows.

{\small 
    \begin{equation} \label{equal:dtc_fd}
    p_{fd} = \begin{cases} 
    1 & \textnormal{if}\  t + c_g + c_b < k \\
    \frac{k}{t + c_g + c_b} \times \frac{k-1}{t + c_g + c_b-1} & \textnormal{otherwise}\ \\
    \end{cases}
    \end{equation}
}

The unbiasedness of DTC-FD means the expected values of the estimation of global and local triangles are equal to the true counts. Due to the space limitation, see \cite{shin2018think} for a detailed proof.
\section{Experiments}\label{sec-expe}

In this section, we present the experimental results of our proposed algorithms and provide a comprehensive comparison with state-of-the-art algorithms.

\subsection{Experimental Settings}

\subsubsection{Platform}

All experiments were conducted on a machine equipped with an Intel(R) Xeon(R) Gold 6148 CPU @ 2.40GHz. Our proposed algorithms for triangle counting were implemented using C++ and MPICH.

\subsubsection{Datasets}All the experiments were performed on the following real-world graph datasets, described in Table \ref{table:dataset}. We removed all parallel, self-loop, and directed edges. 

\begin{table}[t]
    \vspace{-2mm}
    \centering
    \caption{Summary of the real-world graph datasets \label{table:dataset}}
    \begin{tabular}{l|c|c|l}
        \toprule
        {\bf Name} & {\bf \# Nodes} & {\bf \# Edges} & {\bf Description} \\
        \midrule
        \textbf{Arxiv} \cite{gehrke2003overview} & $34,546$ & $420,877$  & Citation network \\
        \textbf{Facebook} \cite{viswanath2009evolution} & $63,731$ & $817,090$ & Social network\\
        \textbf{Dblp} \cite{snapnets} & $317,080$ & $1,049,866$ & Collaboration network \\
        \textbf{NotreDame} \cite{snapnets} & $325,729$ & $1,090,108$ & Web graph\\
        \textbf{BerkStan} \cite{snapnets} & $685,230$ & $6,649,470$ & Web graph\\
        \textbf{Youtube} \cite{mislove2007measurement} & $3,223,589$ & $9,376,594$  & Social network \\
        \textbf{Skitter} \cite{snapnets} & $1,696,415$ & $11,095,298$ & Internet graph \\
        \textbf{LiveJournal} \cite{snapnets} & $3,997,962$ & $34,681,189$ & Social network \\ 
        \bottomrule
    \end{tabular}
\end{table}

\subsubsection{Evaluation Metrics} We evaluated the accuracy and performance of proposed distributed streaming algorithms over real-world datasets by the following measurable metrics.

$\bullet~ \textit{Mean Global Error}$ $\hfill$

Let $x$ be the exact number of global triangles, and $\hat{x}$ be the approximate value estimated by algorithms. Considering $x$ may be equal to 0 in extreme circumstances, we added 1 to both $x$ and $\hat{x}$. Let $n$ denote the number of runs for each experiment. Then, the $\textit{global error}$ was
\begin{equation*}
\frac{1}{n} \sum\nolimits_{i=1}^{n}\frac{|x-\hat{x}_i|}{x+1}
\end{equation*}

$\bullet~ \textit{Mean Local Error}$ $\hfill$

Likewise, we let $x_u$ be the exact local counts of node $u$ in $\mathcal{G}$, and $\hat{x}_{ui}$ be the estimated one for $i$-$th$ run, as each measure was performed by multiple times. Then, the $\textit{local error}$ was 
\begin{equation*}
\frac{1}{n} \sum\nolimits_{i=1}^{n} \left\{
\frac{1}{|\mathcal{V}|} \sum\nolimits_{u \in {\mathcal{V}}}
\frac{|x_u - \hat{x}_{ui}|}{x_u + 1} \right\}
\end{equation*}

$\bullet~ \textit{Global variance}$ $\hfill$

The $\textit{global variance}$ of estimated global triangle counting $\hat{x}$ was defined as follows.

\begin{equation*}
\frac{1}{n} \sum\nolimits_{i=1}^{n}{(x - \hat{x}_i)}^2
\end{equation*}

$\bullet~ \textit{Pearson Correlation Coefficient}$  $\hfill$

This showed the closeness of relationship between two variables $x$ and $y$ \cite{lim2015mascot}, expressed by a linear function. Given $x$ and $\hat{x}$ as the exact and estimated counts for triangles respectively, the definition was illustrated as follows.

\begin{equation*}
\rho(x,\hat{x}) = \frac{Cov(x,\hat{x})}{\sigma_x \sigma_{\hat{x}}}
\end{equation*}

\begin{figure*}
    \centering
    \includegraphics[width=0.5\linewidth]{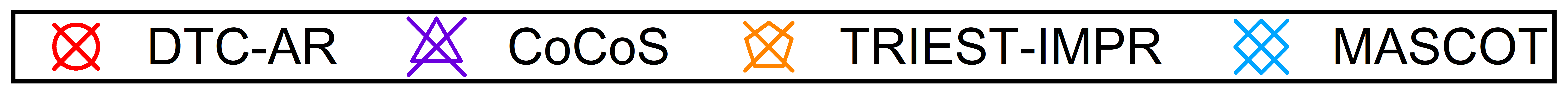}\\
    \vspace{0.5mm}
    
    \subfigure[Youtube]{
        \includegraphics[width= 0.2\linewidth]{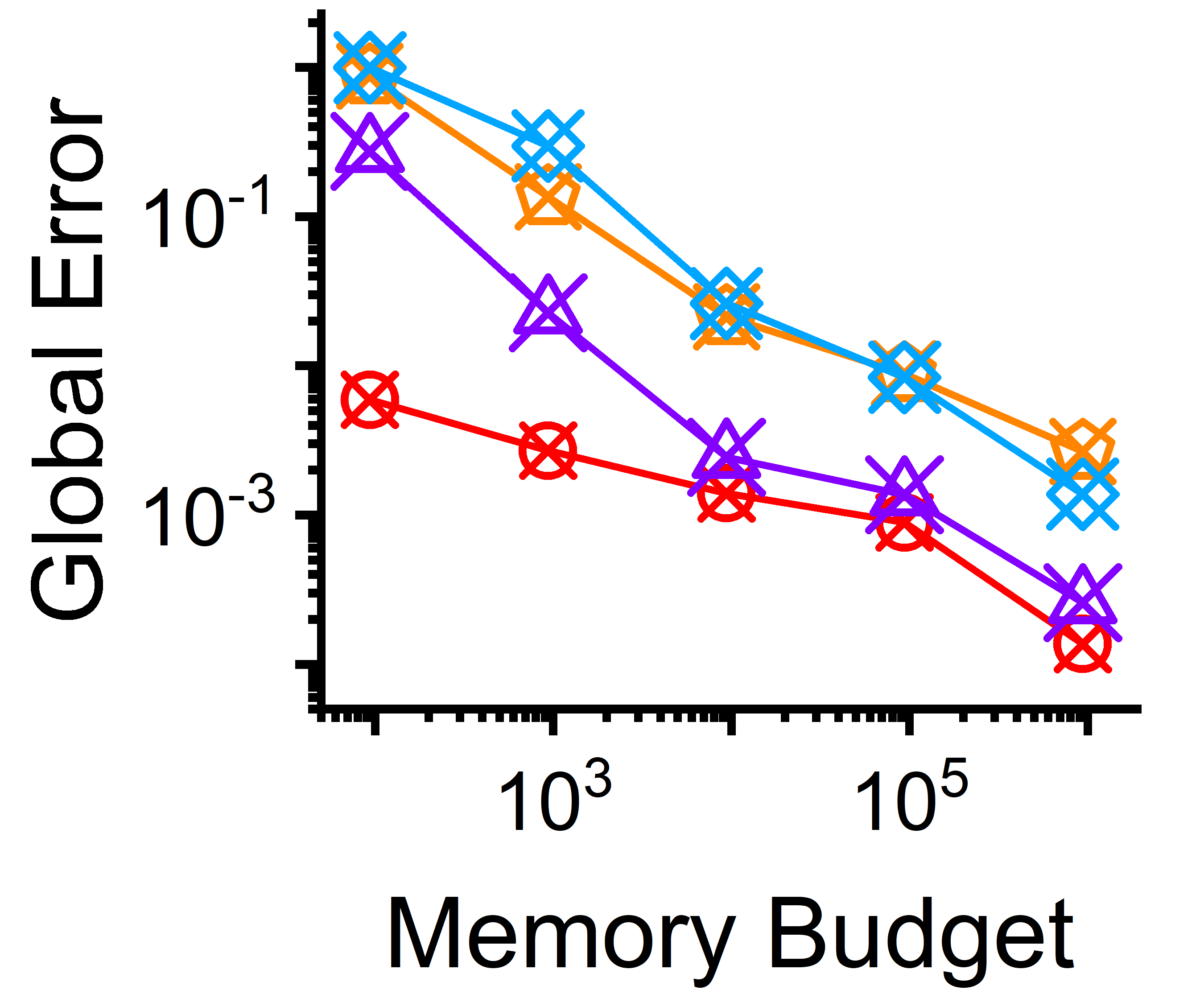}
    }
    \subfigure[Skitter]{
        \includegraphics[width= 0.2\linewidth]{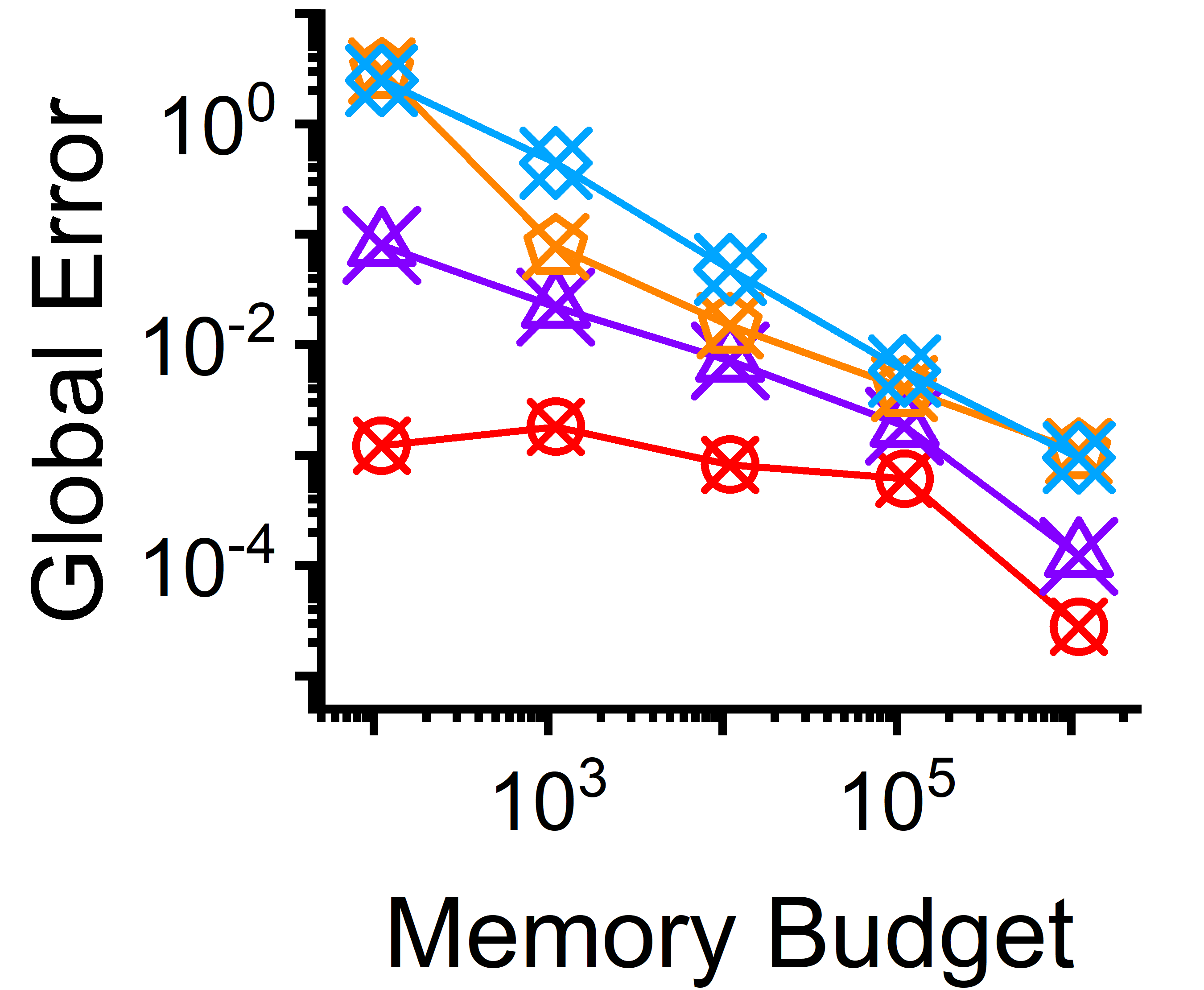}
    }
    \subfigure[Youtube]{
        \includegraphics[width= 0.2\linewidth]{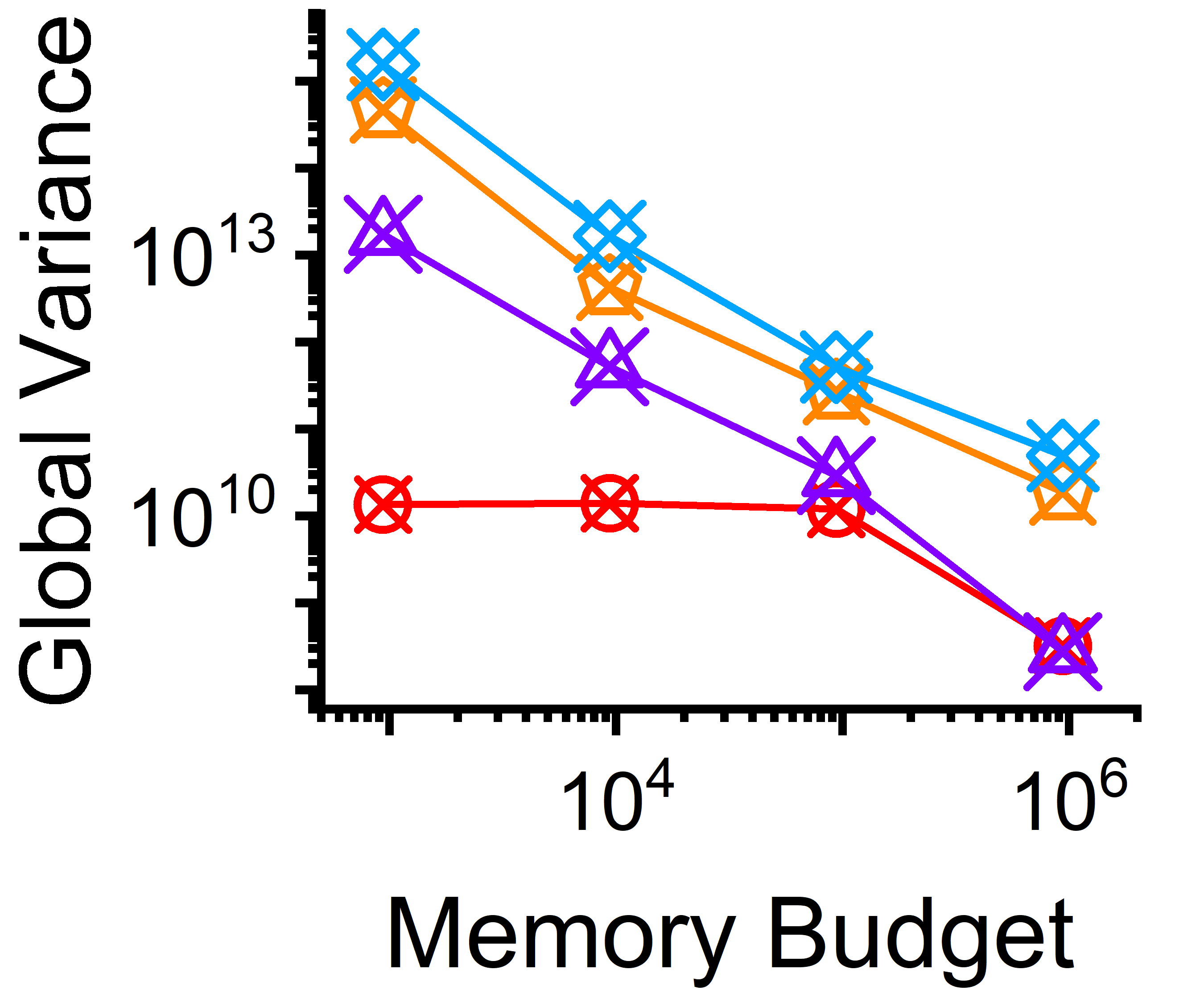}
    }
    \subfigure[Skitter]{
        \includegraphics[width= 0.2\linewidth]{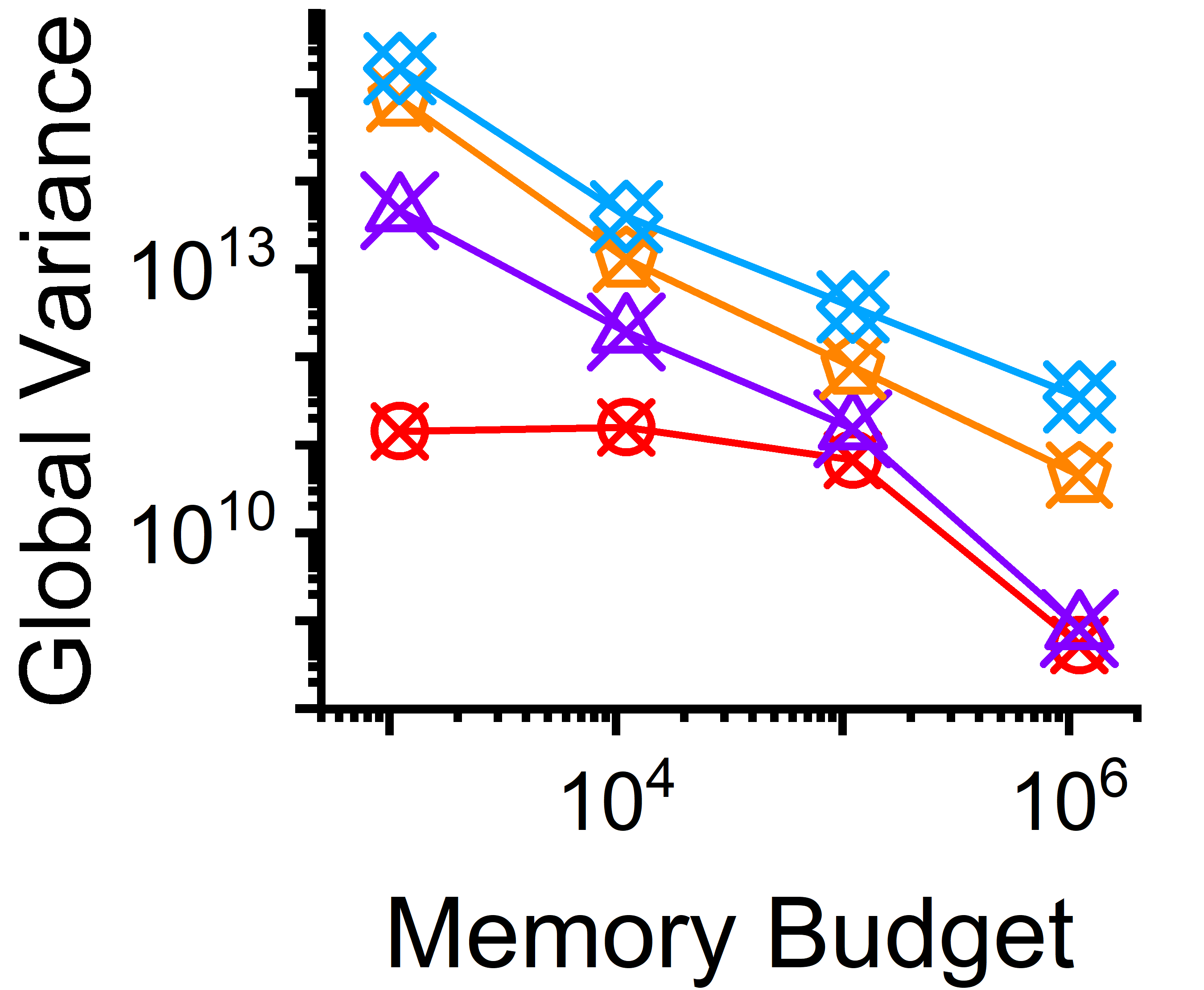}
    }
    \\
    
    \subfigure[Youtube]{
        \includegraphics[width= 0.2\linewidth]{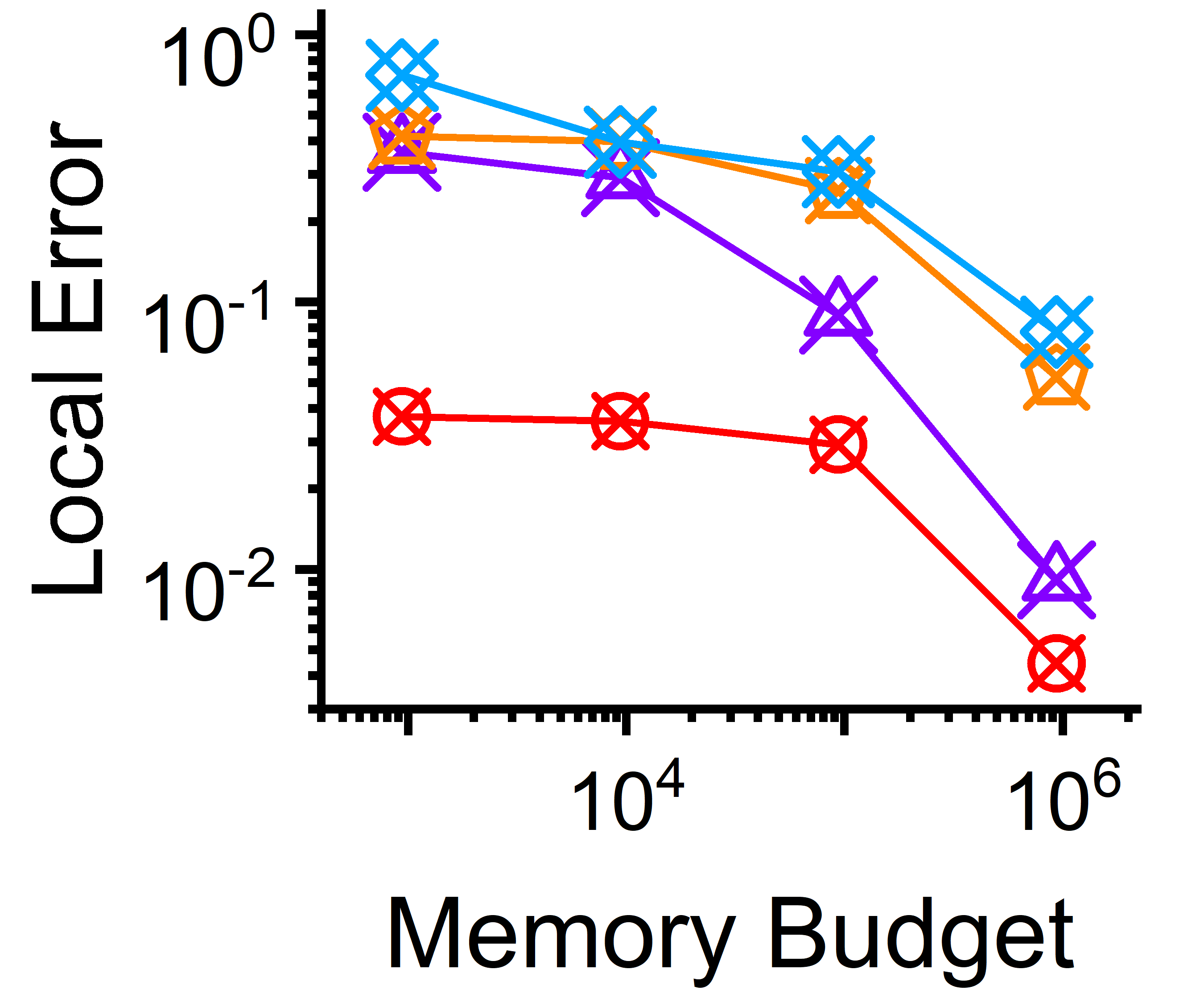}
    }
    \subfigure[Skitter]{
        \includegraphics[width= 0.2\linewidth]{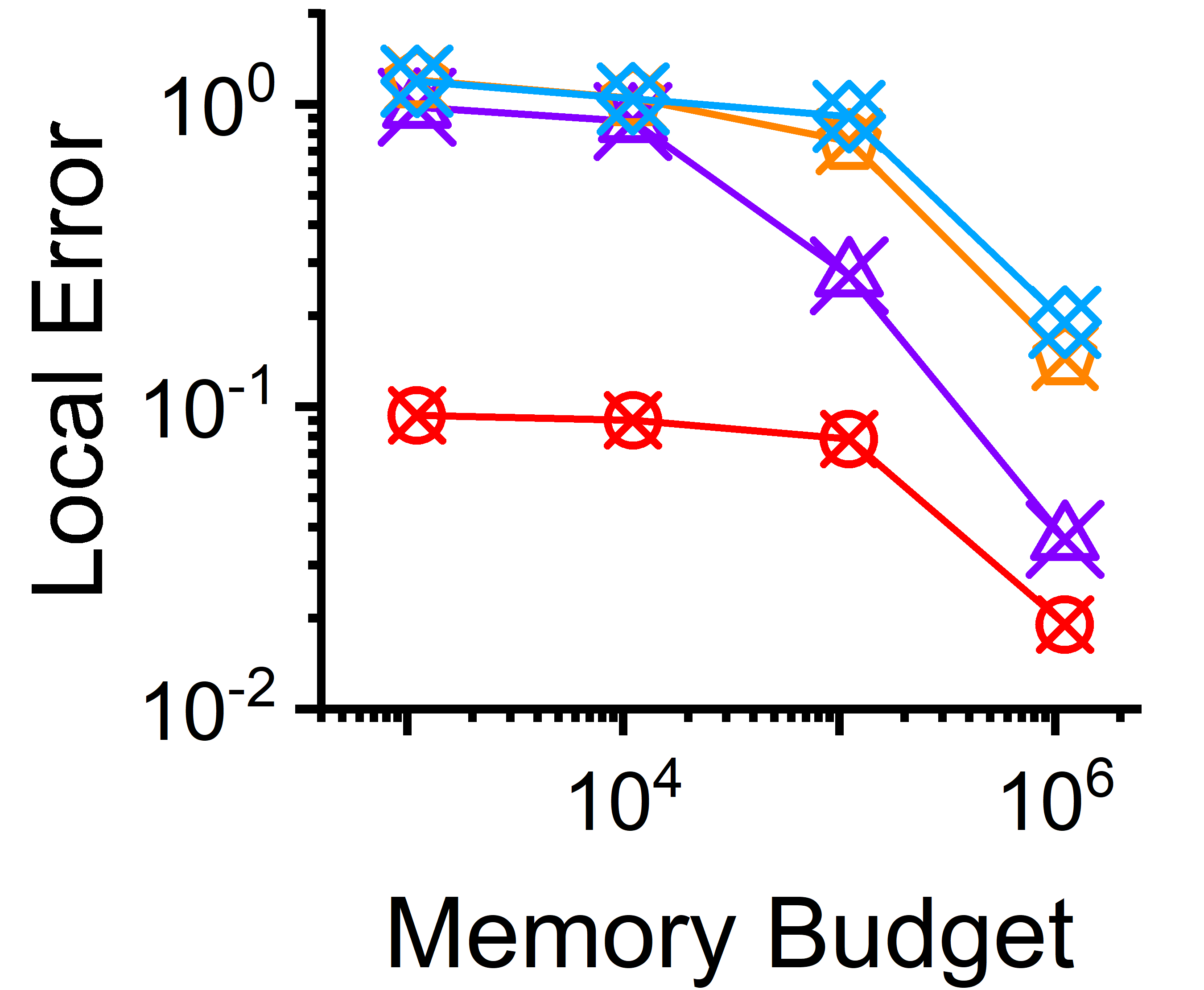}
    }
    \subfigure[Youtube]{
        \includegraphics[width= 0.2\linewidth]{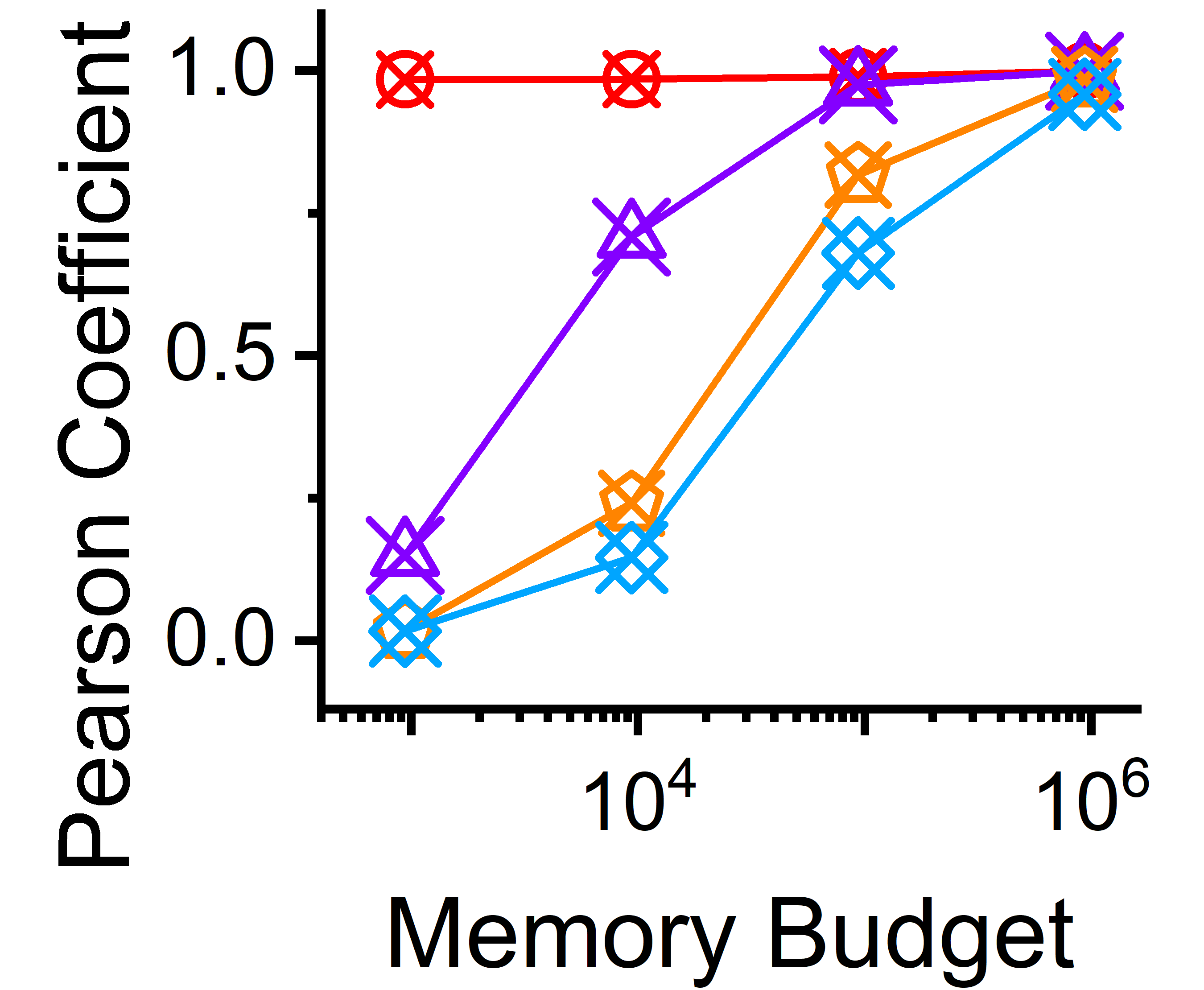}
    }
    \subfigure[Skitter]{
        \includegraphics[width= 0.2\linewidth]{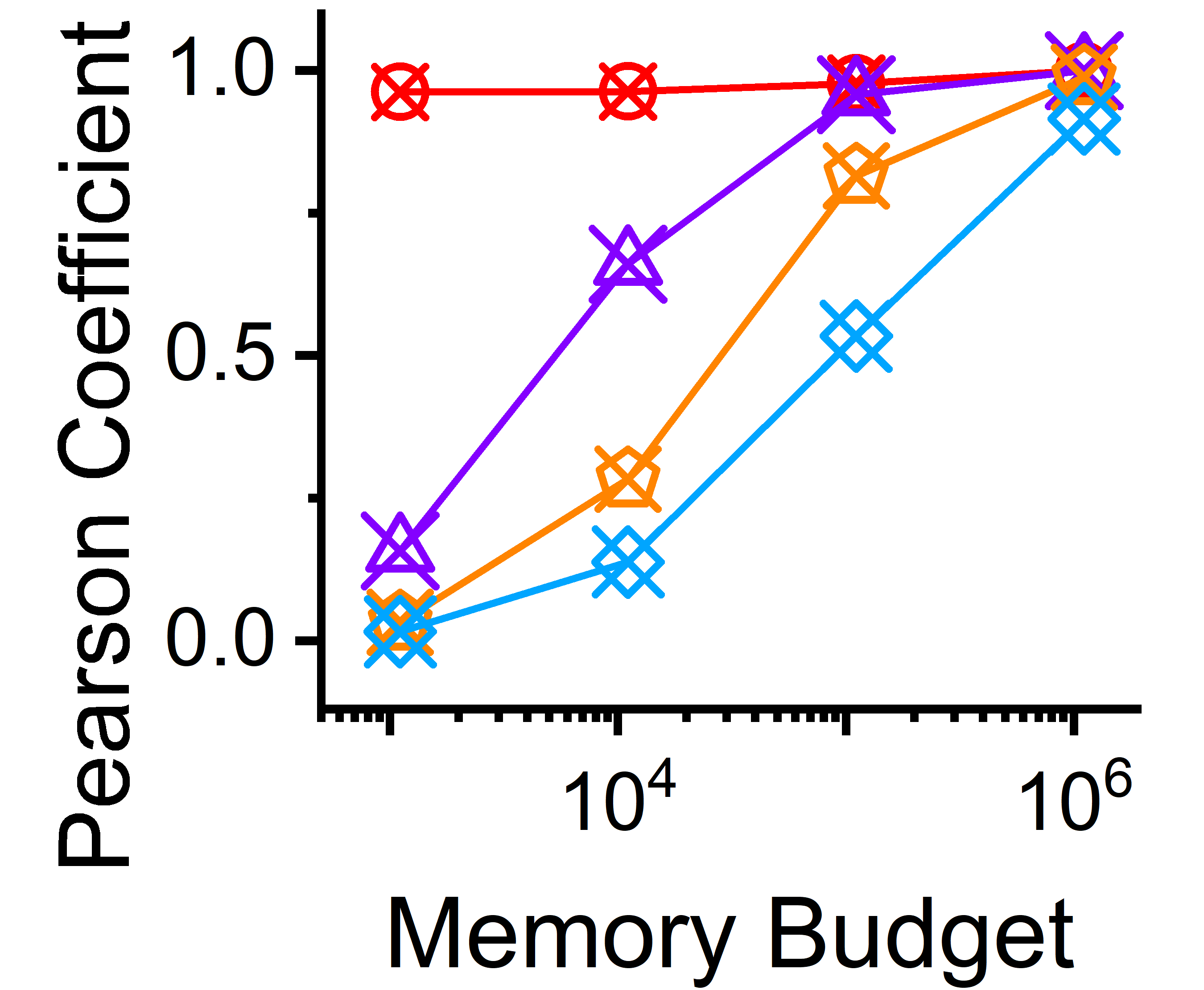}
    } 
    \\

    \subfigure[Dblp]{
        \includegraphics[width= 0.2\linewidth]{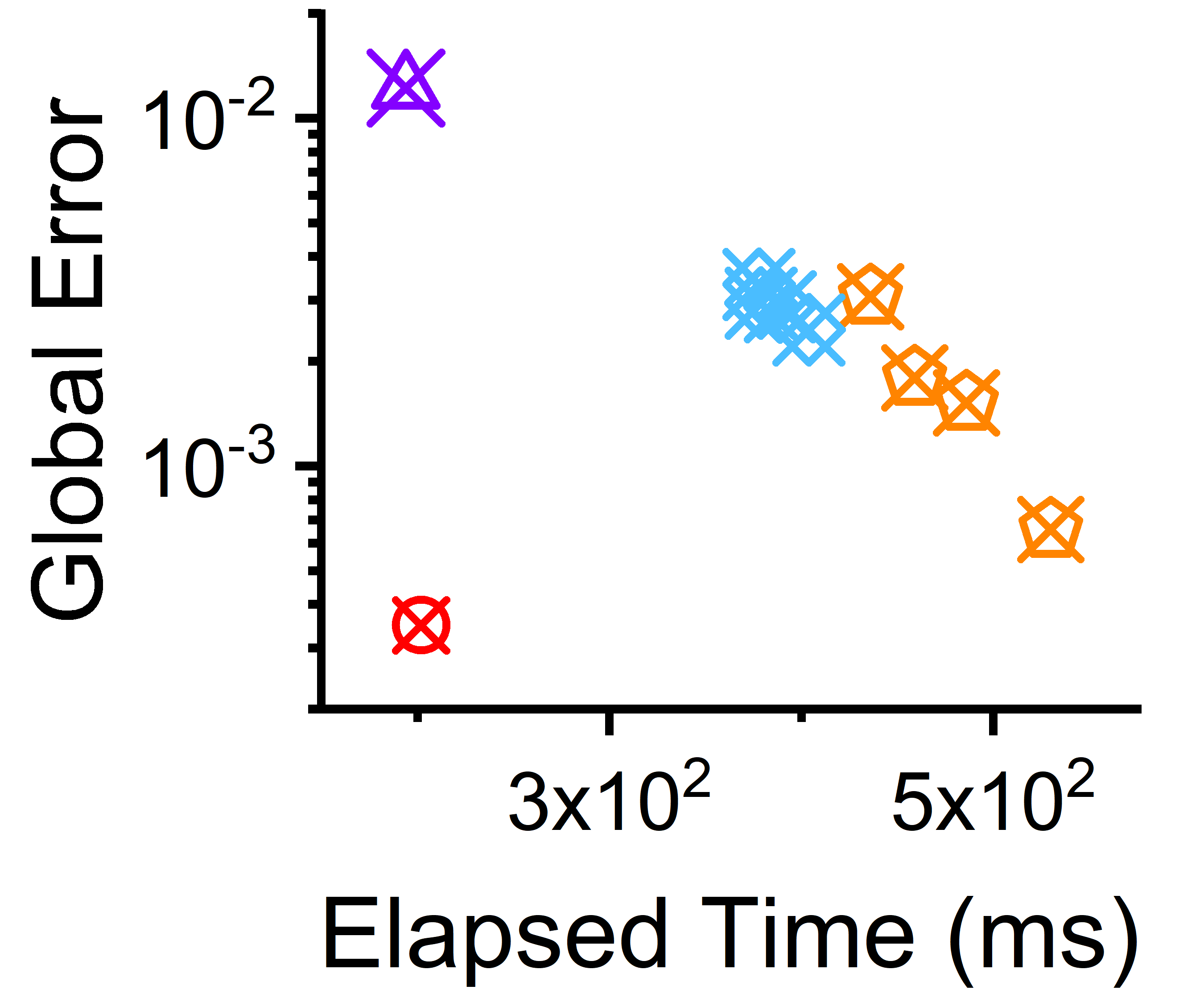}
    }
    \subfigure[Dblp]{
        \includegraphics[width= 0.2\linewidth]{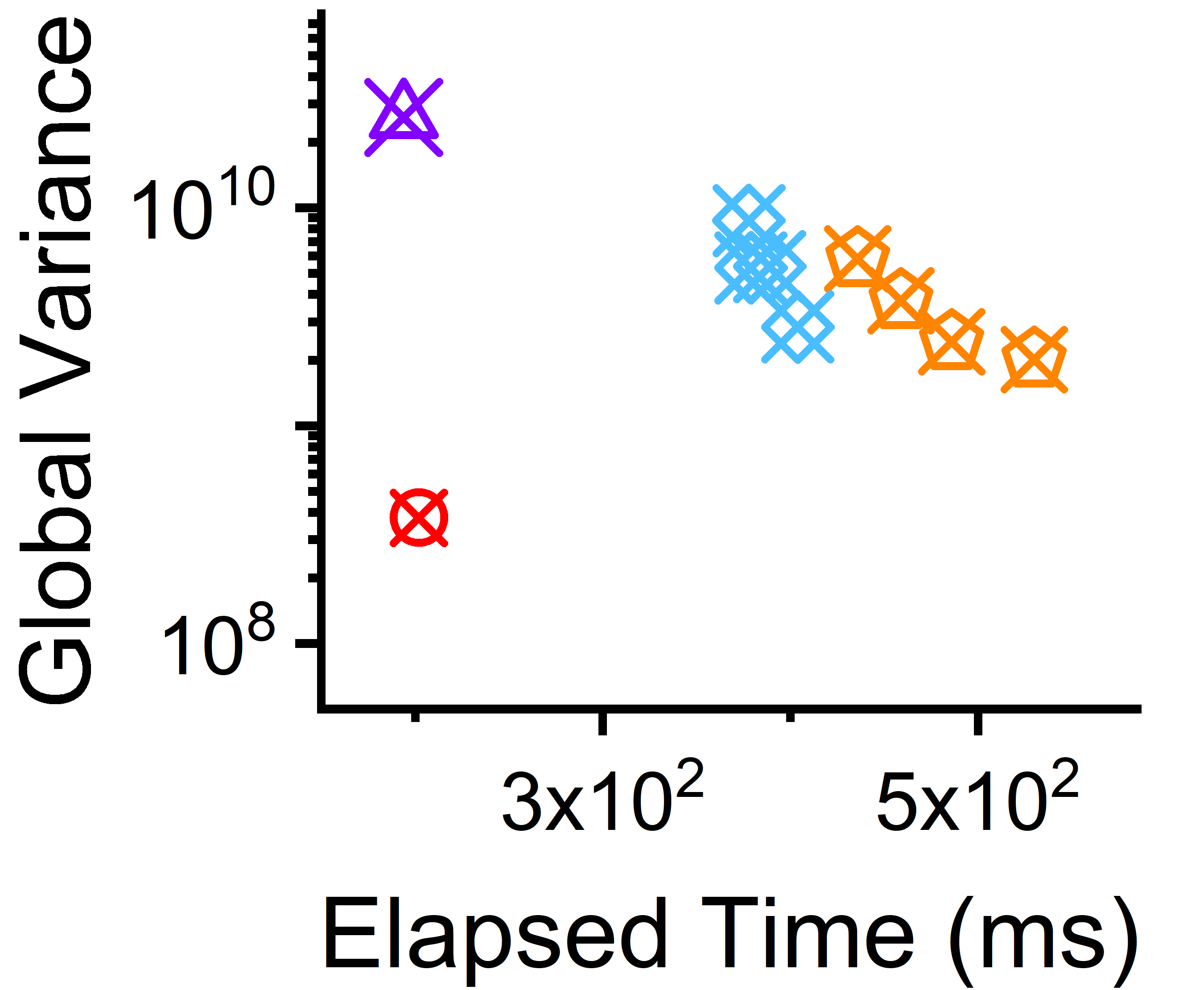}
    }
    \subfigure[Dblp]{
        \includegraphics[width= 0.2\linewidth]{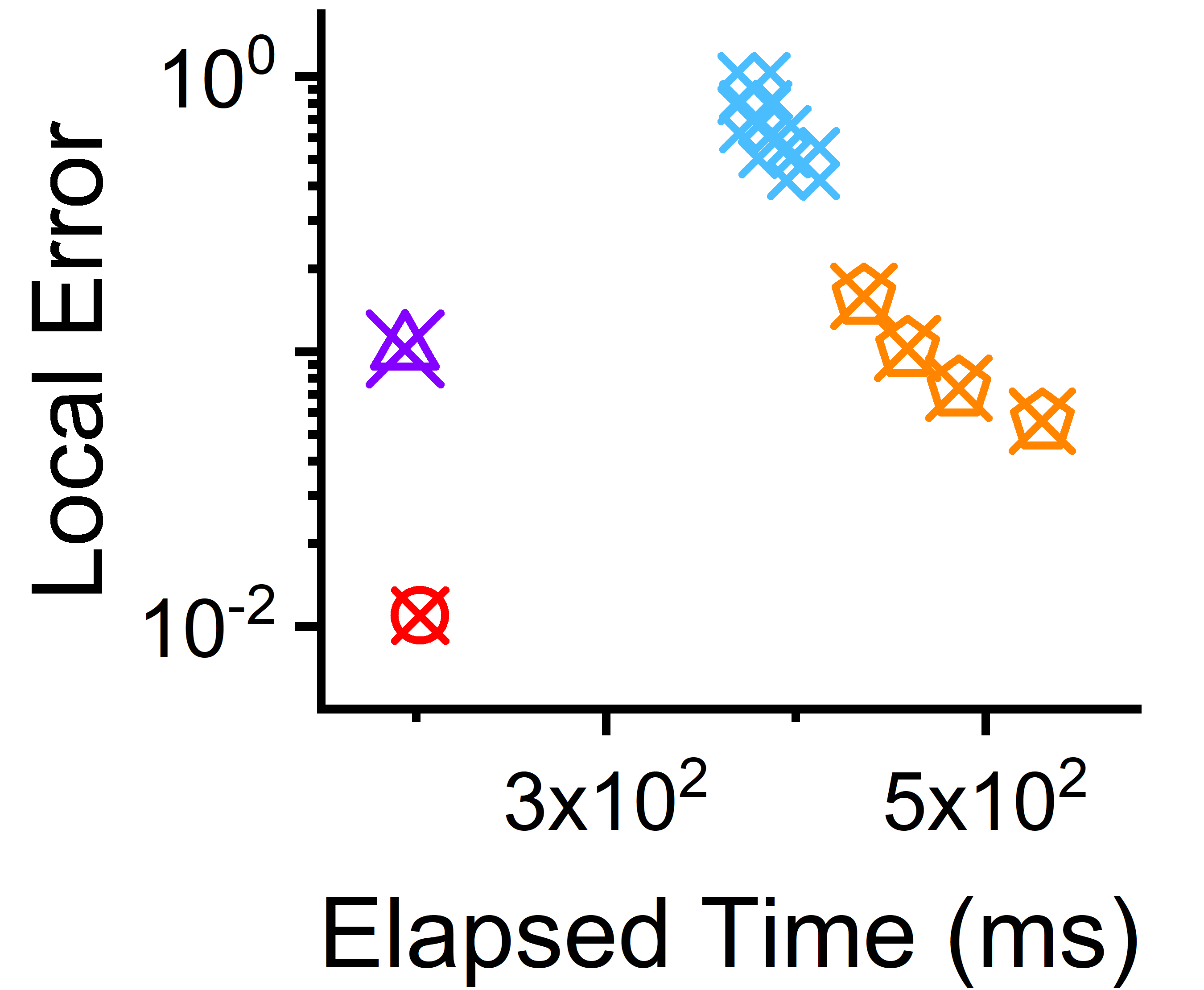}
    }
    \subfigure[Dblp]{
        \includegraphics[width= 0.2\linewidth]{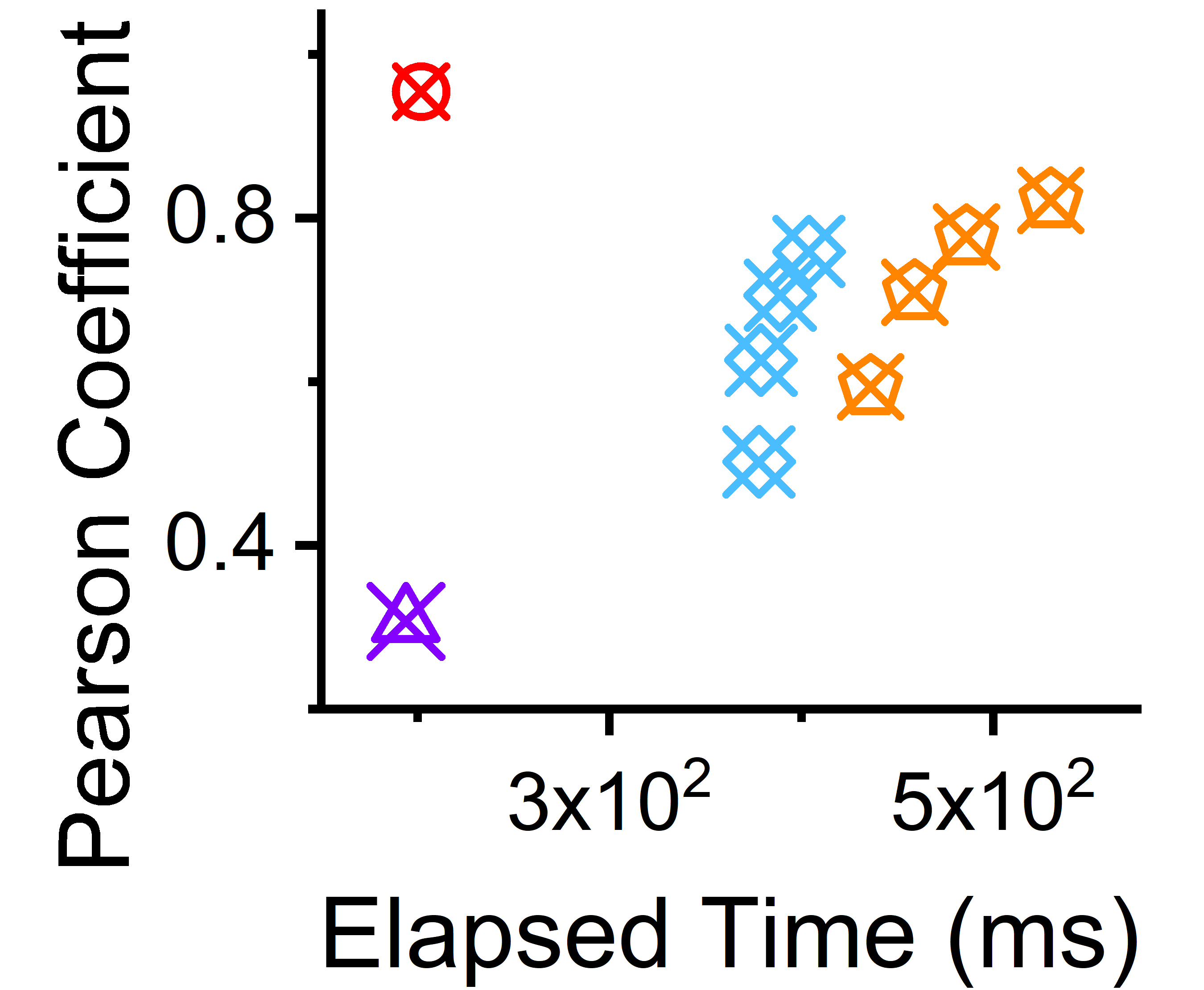}
    } 
    \\
    \vspace{-2mm}
    \caption[Accuracy and Speed of DTC-AR]{\label{fig:comrarison:dtc-ar:accuracy} \figsummary{Accuracy and Speed of DTC-AR.}
        DTC-AR demonstrates significantly enhanced accuracy and speed compared to other algorithms. It achieves up to 2029.4$\times$ and 27.1$\times$ higher accuracy than MASCOT in terms of global error and local error, respectively. Moreover, the Pearson coefficient of DTC-AR consistently approaches 1, indicating its superior performance compared to alternative algorithms.
    }
\end{figure*}

\subsubsection{Implemented Algorithms} $\hfill$

$\bullet~ $ {Distributed algorithms for triangle counting}

\begin{itemize}
    \item [-] CoCoS \cite{shin2021cocos}: state-of-the-art distributed streaming algorithms.
    \item [-] DTC-AR: our proposed distributed streaming algorithm by adaptive resampling.
    \item [-] DTC-FD: our proposed distributed streaming algorithm for handling fully dynamic graph streams.
\end{itemize}

$\bullet~ $ {Single-machine algorithms for triangle counting}

\begin{itemize}
    \item [-] TRIEST-IMPR\cite{stefani2017triest}, MASCOT\cite{lim2015mascot}: only for insertion edge in graph streams.
    \item [-] ThinkDAcc\cite{shin2018think}, TRIEST-FD \cite{stefani2017triest}: support fully dynamic graph streams.
    \item [-] MASCOT-FD: adapted from MASCOT to handle fully dynamic graph streams.
\end{itemize}

Among above state-of-the-art streaming algorithms for triangle counting, we adapted MASCOT-FD from MASCOT to handle both edge insertion and edge deletion. They all achieved better accuracy than other well-known triangle counting algorithms. In order to illustrate the advantages, we compared our proposed distributed streaming algorithms with them by various evaluation metrics. We measured each independent experiment by 100 times, and calculated the average to guarantee the statistical stability of assessment. In distributed settings, we chose samping threshold $\mathcal{R}$ of DTC-AR as $0.2$, if not otherwise stated.

\subsection{DTC-AR for Insertion-only Streams}

To the best of our knowledge, DTC-AR is the first distributed streaming algorithm for both global and local triangle counting by adaptive resampling in evolving graph streams. Since the size of graph streams is enormous and usually increases as time flies in various real-world application scenarios, it is almost impracticable to set the appropriate parameters for existing streaming algorithms. However, state-of-the-art Reservoir-based and Bernoulli-based sampling algorithms all require prior knowledge about the size of graph streams for the best trade-off between accuracy and storage space.

\subsubsection{Accuracy}

For analyzing the accuracy for both global and local triangle counting, we measured the mean global error, global variance, mean local error and pearson correlation with different storage budgets. For expressing the uncertainty of the size of graph streams, we chose a large range of variation for the storage budget (i.e., \{0.00001, 0.0001, 0.001, 0.01, 0.1\} relative to the size of edges in each graph datasets listed in Table \ref{table:dataset}). The number of workers was 10 for both distributed streaming algorithms DTC-AR and CoCoS. As shown in Figure \ref{fig:comrarison:dtc-ar:accuracy}, DTC-AR always achieved higher accuracy than the baselines. Specifically, DTC-AR was up to 2029.4$\times$ and 27.1$\times$ smaller estimate error than MASCOT in terms of global error and local error, respectively. Moreover, DTC-AR maintained high accuracy by adaptively increasing storage budget to guarantee a specified sampling threshold in evolving graph streams, even though the initial storage budget was minuscule.

\subsubsection{Speed}

We measured both the speed and accuracy of distributed streaming algorithms (proposed DTC-AR, CoCoS) and state-of-the-art single-machine algorithms (TRIEST-IMPR, MASCOT). The storage budget of TRIEST-IMPR was from 2\% to 5\% of the number of edges, and the sampling probability of MASCOT is from 2\% to 5\%. For purposes of comparison, we only stored 0.1\% of graph dataset Dblp in distributed settings. And we set the number of workers be 10 for DTC-AR and CoCoS. In Figure \ref{fig:comrarison:dtc-ar:accuracy} (i-l), we provided the evaluation metrics and elapsed time averaged over 100 times running on graph dataset Dblp. DTC-AR yielded best trade-off between speed and accuracy, as shown in Figure \ref{fig:comrarison:dtc-ar:accuracy} (i-l). Particularly, DTC-AR achieved up to 35.0$\times$ more accuracy than CoCoS in terms of global error, when they consumed similar time. Since DTC-AR utilized more memory space to store sampled edges by adaptive sampling method, it actually consumed a slightly higher time than CoCoS. Furthermore, DTC-AR consistently outperformed TRIEST-IMPR and MASCOT in terms of speeds and accuracies.

\begin{figure}
    \centering
    \vspace{0.5mm}
    \hspace{-4mm}
    \vspace{0.5mm}
    \subfigure[BerkStan]{
        \includegraphics[width= 0.4\linewidth]{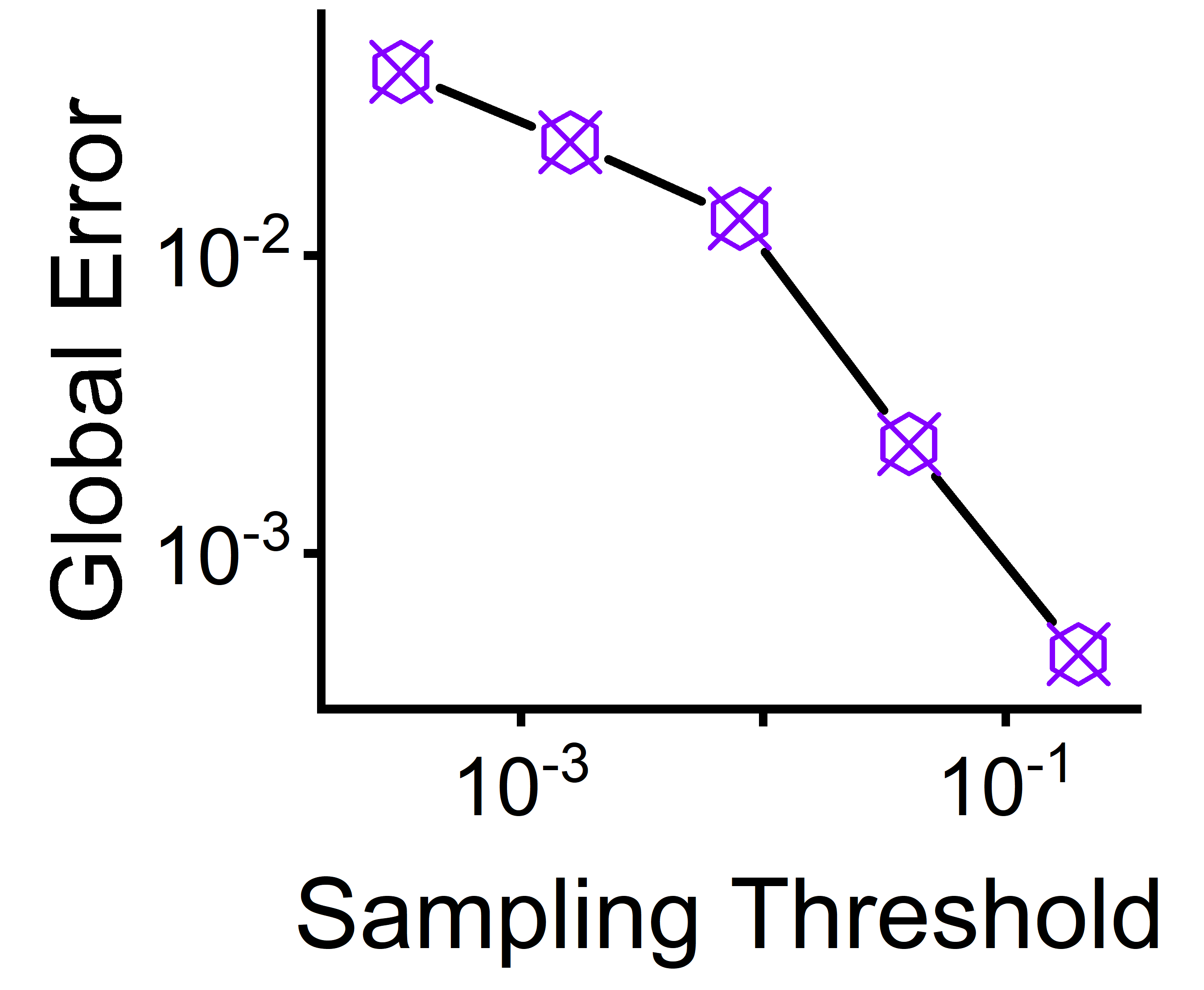}
    }
    \subfigure[Youtube]{
        \includegraphics[width= 0.4\linewidth]{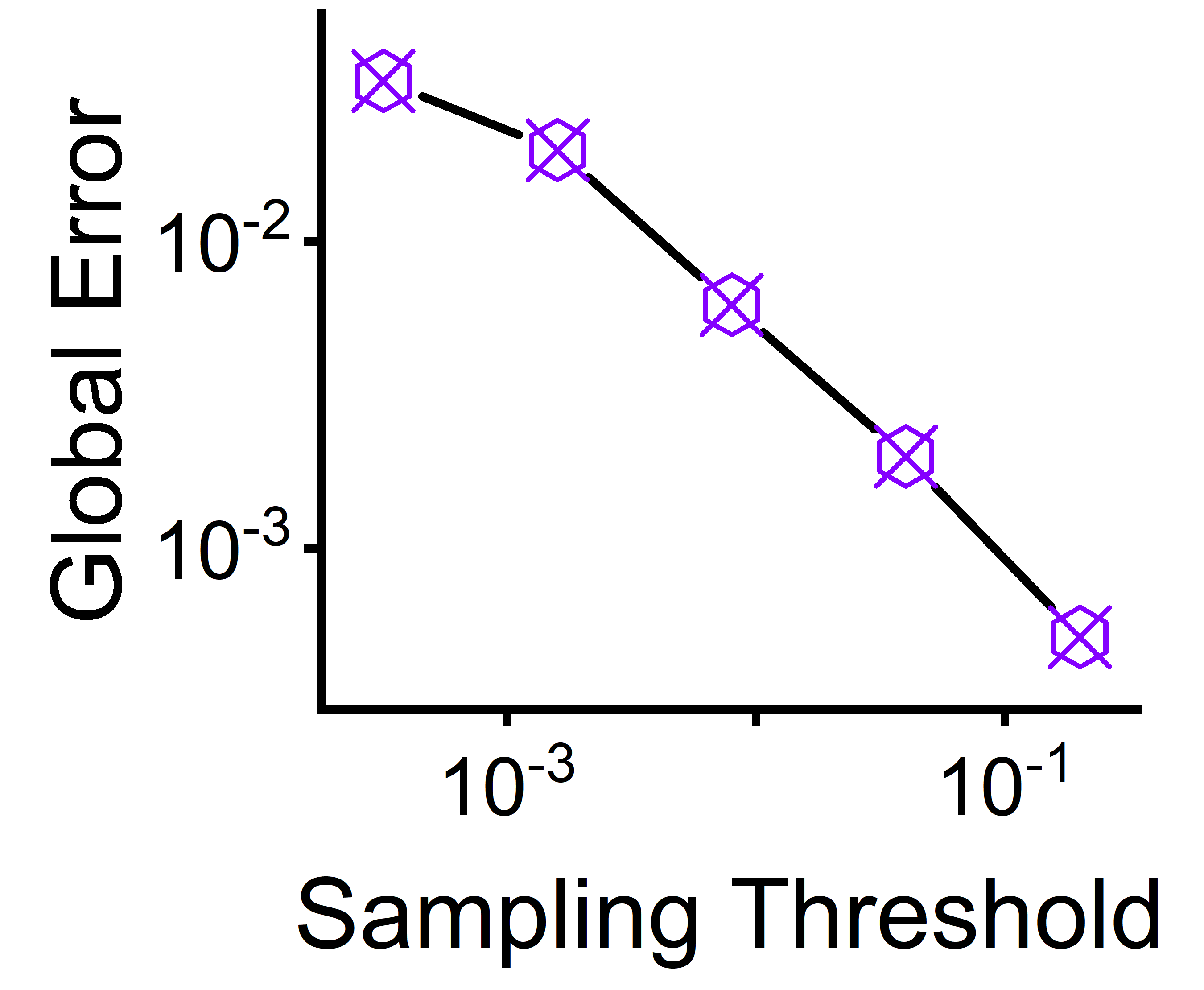}
    }
    \\
    \vspace{-2mm}
    \caption[Scalability of DTC-AR]{\figsummary{Scalability of DTC-AR.} The scalability of DTC-AR is directly proportional to the sampling threshold in graph streams, ensuring efficient performance as the threshold increases.
    }
    \label{fig:comparison:dtc_ar_multi_threshold}
\end{figure}

\subsubsection{Scalability}

We measured how different sampling thresholds affected the accuracy of triangle counting of DTC-AR in Figure \ref{fig:comparison:dtc_ar_multi_threshold}. We used 20 workers with storage budget set to $3\times10^4$, and changed the sampling threshold of each dataset from 0.032\% to 20\% to calculate the global errors at the end of each graph streams. On the one hand, Figure \ref{fig:comrarison:dtc-ar:accuracy} (a-b) showed DTC-AR always maintained lower global error than others. Although given different initial storage budgets, DTC-AR achieved almost same accuracy for the given sampling threshold $\mathcal{R}$. On the other hand, Figure \ref{fig:comparison:dtc_ar_multi_threshold} further illustrated that the sampling threshold determined the accuracy for triangle counting in evolving graph streams. And the global error of our proposed DTC-AR reduced linearly with the increase of the sampling threshold. 

In view of unknowing any prior knowledge about the size of real-world graph streams,
setting an appropriate sampling threshold rather than initial memory budget, is practicable and operable. Based on this strategy, DTC-AR enables to keep this scalability for triangle counting in graph streams, which is unsolvable for any other distributed streaming algorithms.

\begin{figure}[t]
    \centering
    \includegraphics[width=0.7\linewidth]{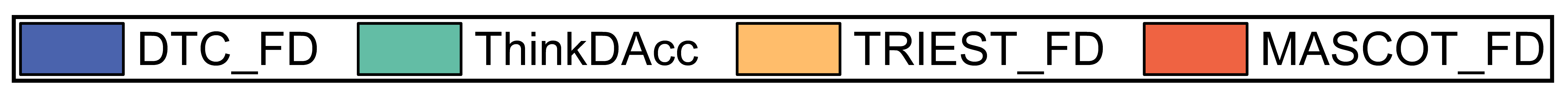}\\
    \vspace{0.5mm}
    \hspace{-4mm}
    \vspace{0.5mm}
    \subfigure[Global Error]{\includegraphics[width= 0.46\linewidth]{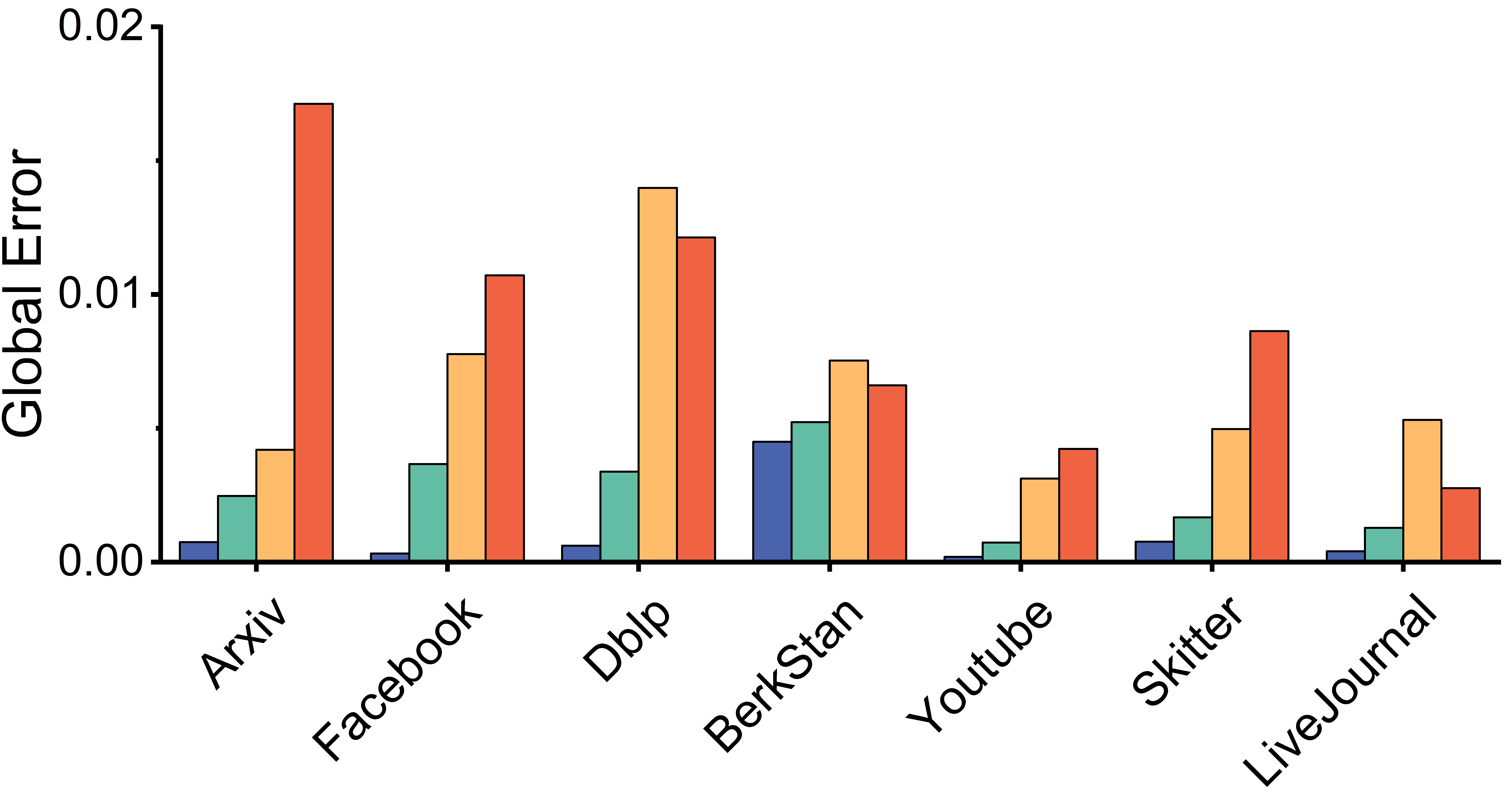}}
    \subfigure[Global Variance]{\includegraphics[width= 0.46\linewidth]{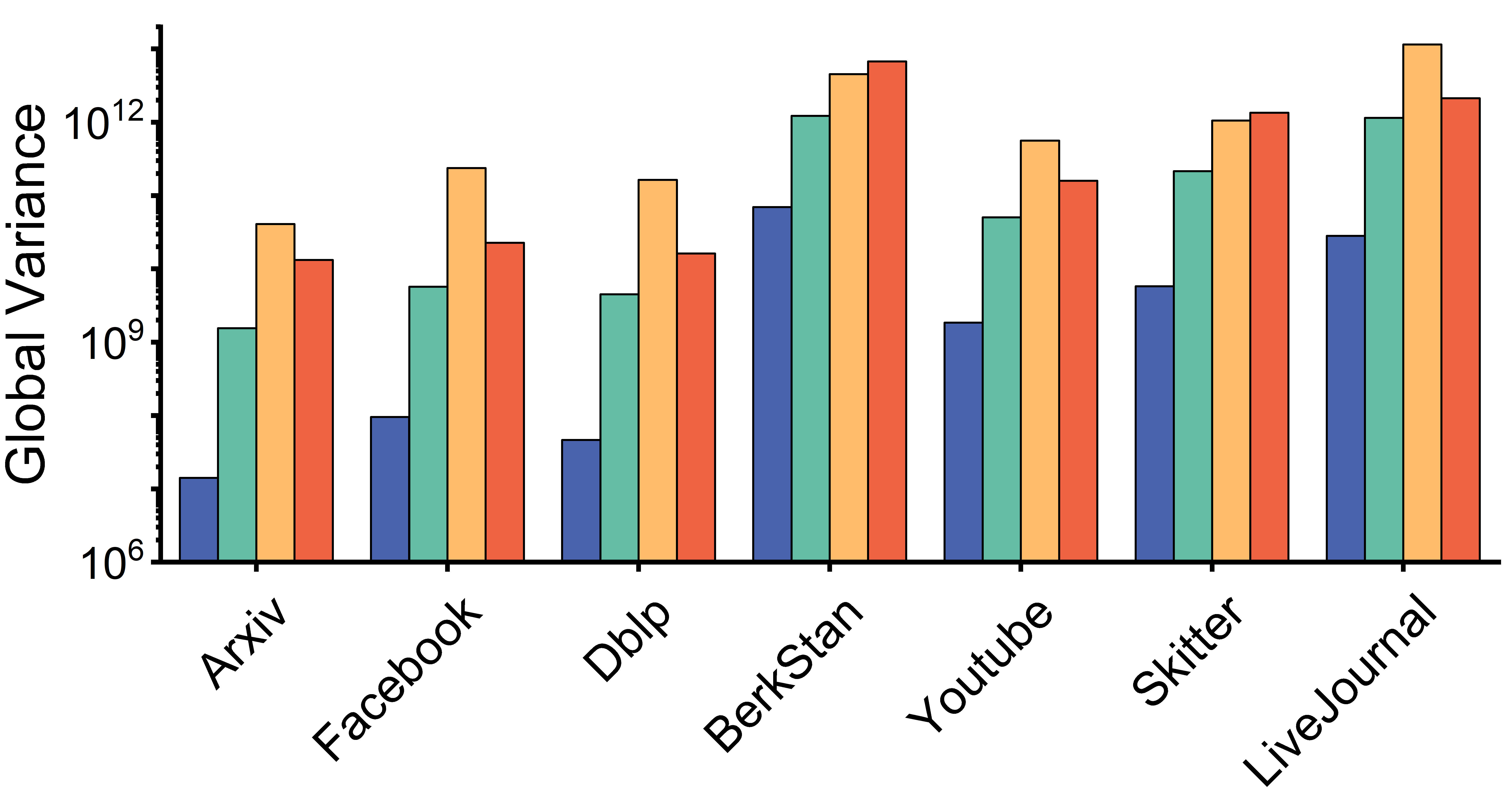}}
    \\
    \subfigure[Local Error]{\includegraphics[width= 0.46\linewidth]{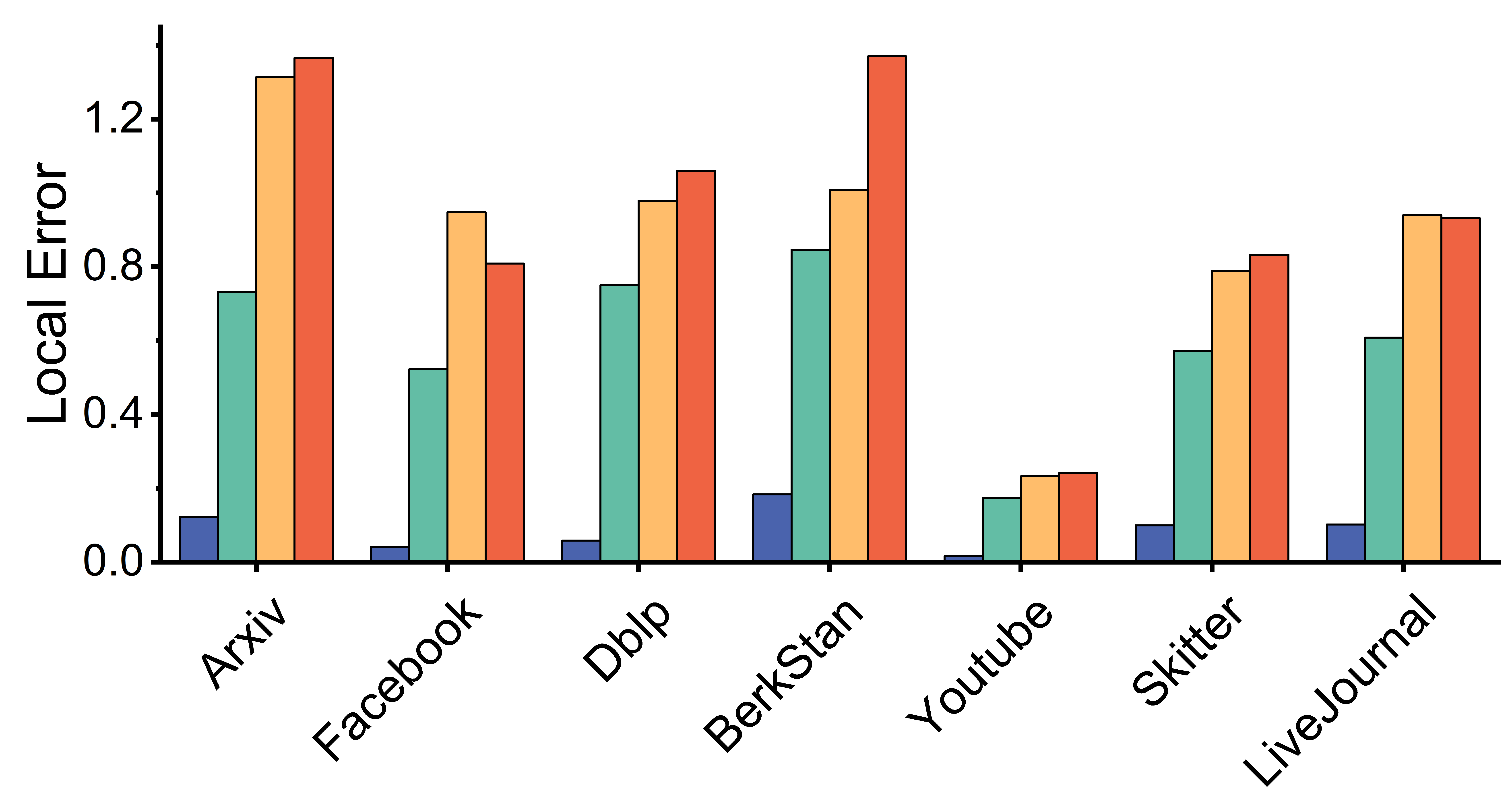}}
    \subfigure[Pearson Coefficient]{\includegraphics[width= 0.46\linewidth]{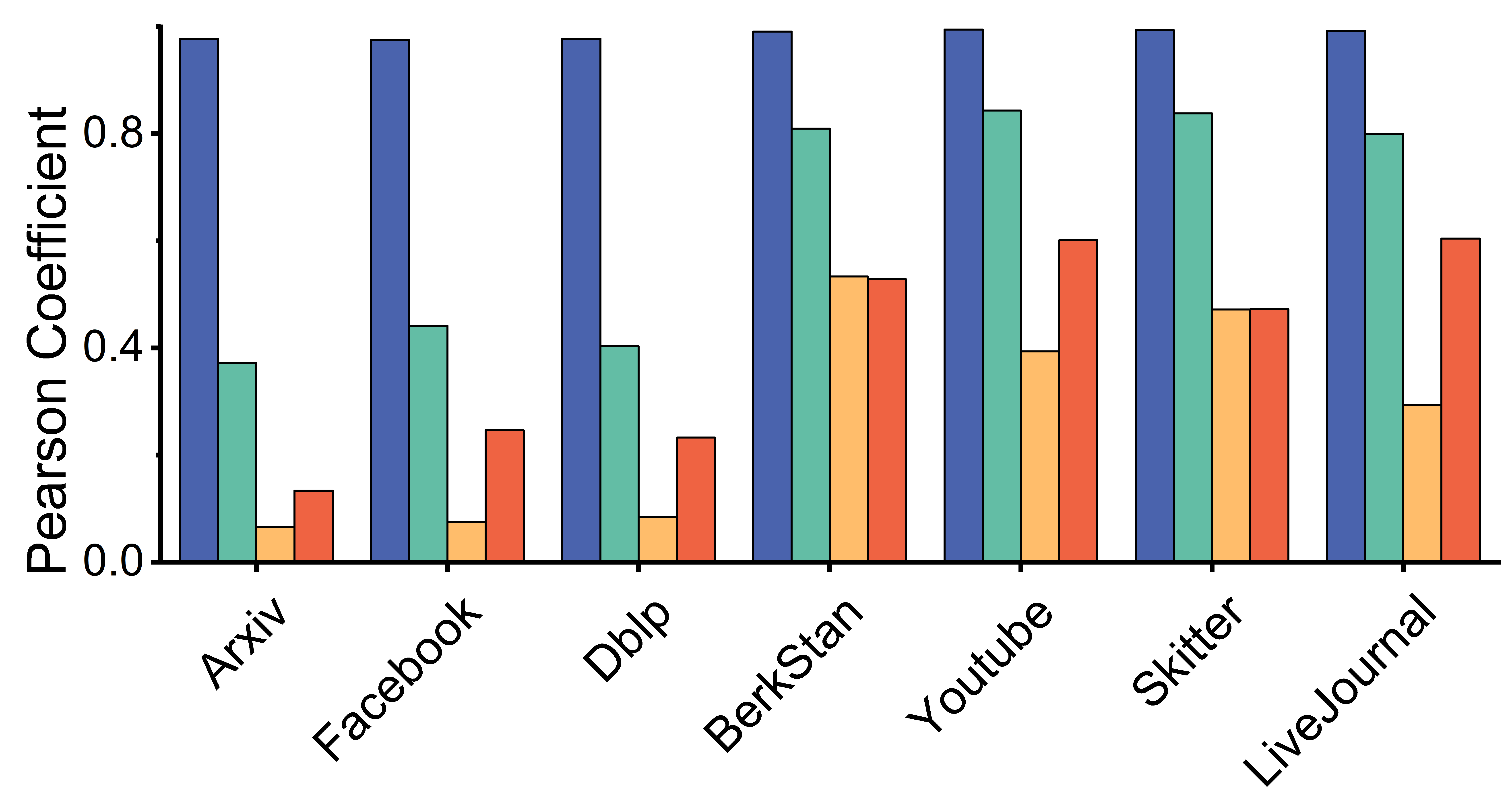}}
    \\
    \vspace{-2mm}
    \caption[Accuracy of DTC-FD]{\figsummary{Accuracy of DTC-FD.}
        DTC-FD demonstrates significantly higher accuracy compared to MASCOT-FD, achieving up to 32.5$\times$ and 19.3$\times$ improvement in global error and local error, respectively. Additionally, DTC-FD outperforms ThinkDAcc with up to 109.7$\times$ improvement in global variance and 2.6$\times$ improvement in the Pearson coefficient.
    }
    \label{fig:fully_dynamic:compare}
\end{figure}

\begin{figure*}
    \centering
    \includegraphics[width=0.5\linewidth]{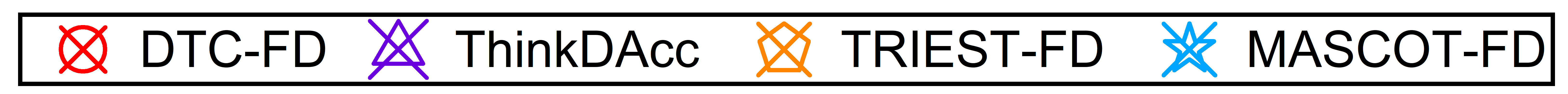}\\
    \vspace{0.5mm}
    
    \subfigure[Dblp]{\includegraphics[width= 0.2\linewidth]{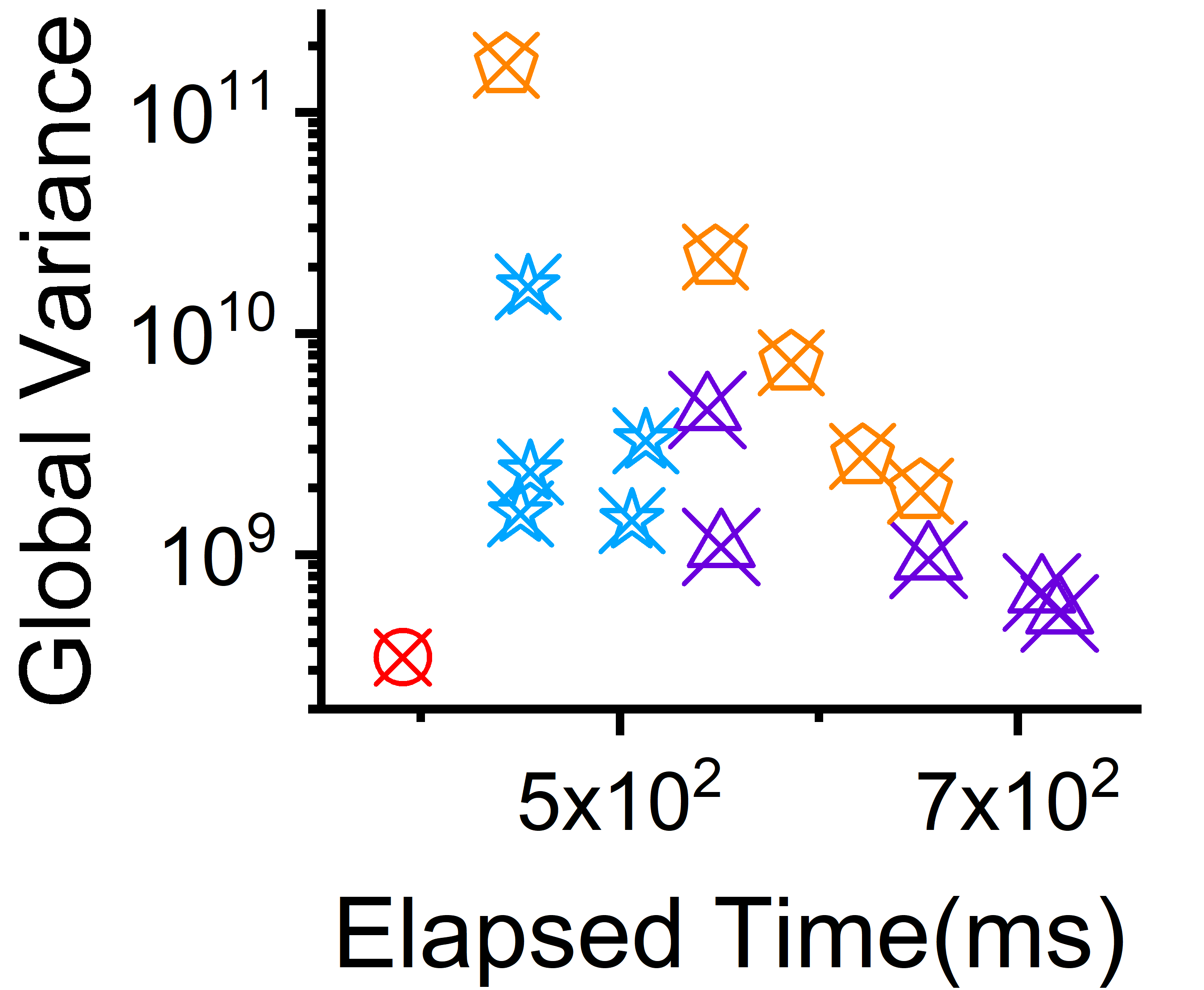}}
    \subfigure[NotreDame]{\includegraphics[width= 0.2\linewidth]{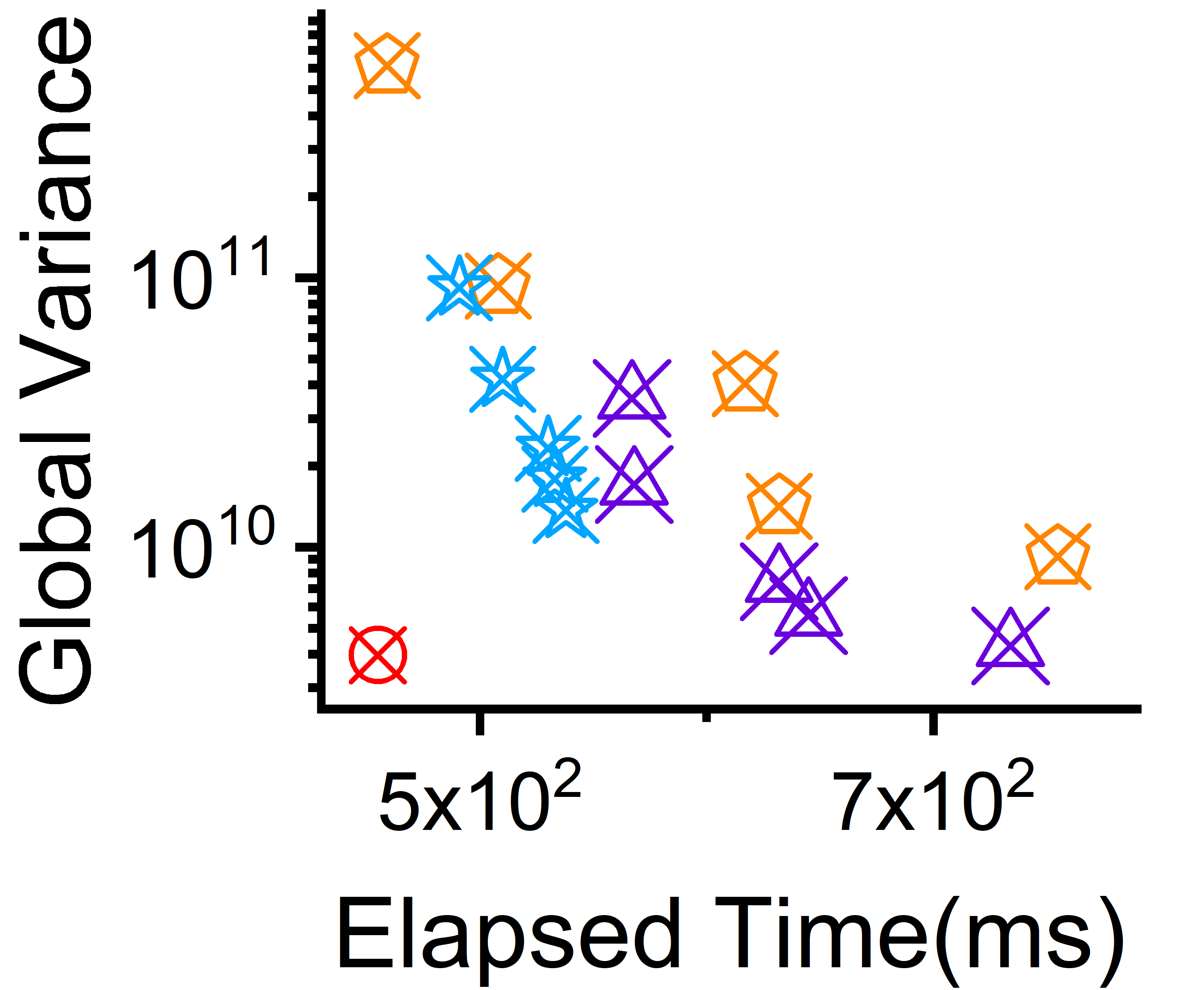}}
    \subfigure[Dblp]{\includegraphics[width= 0.2\linewidth]{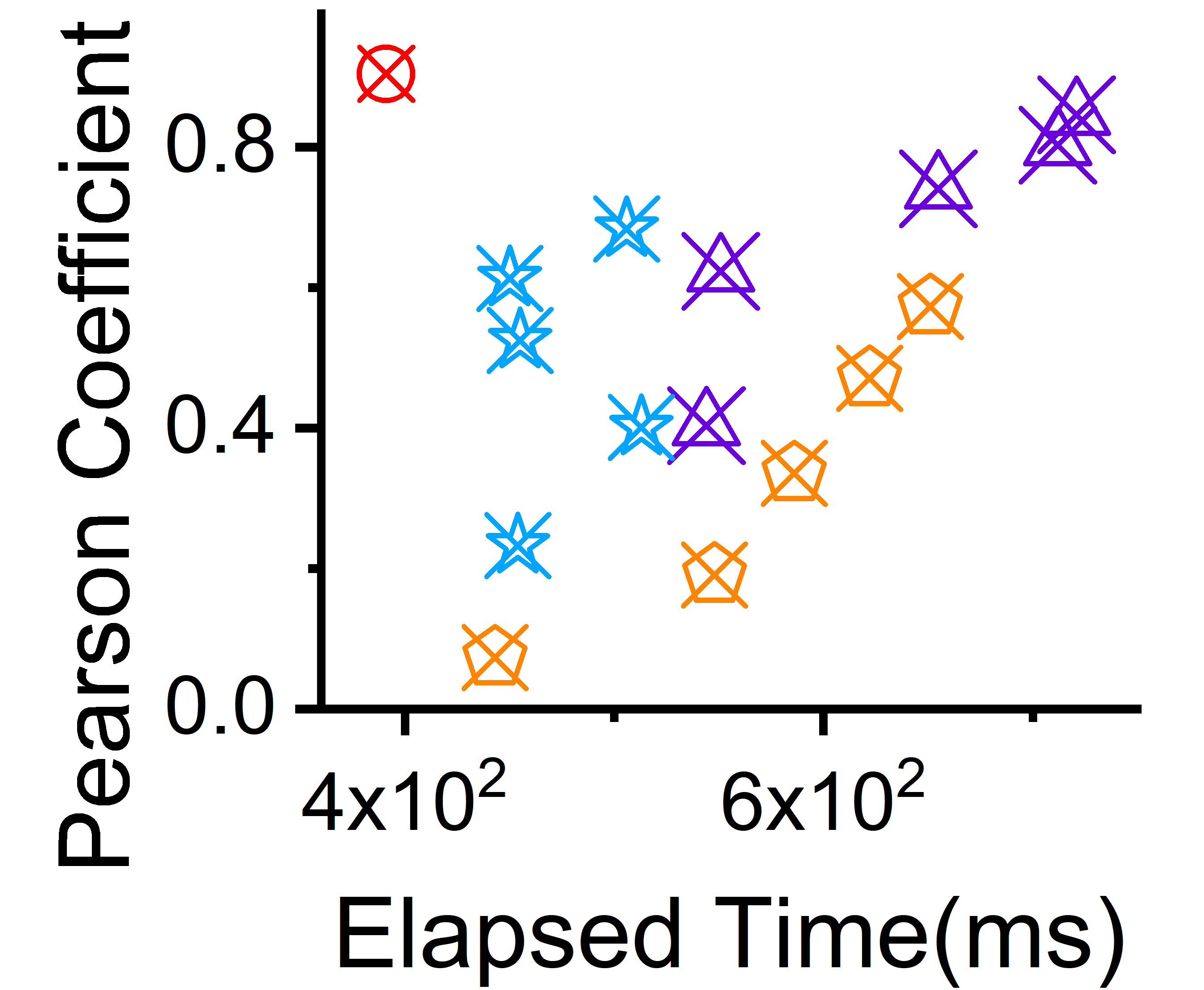}}
    \subfigure[NotreDame]{\includegraphics[width= 0.2\linewidth]{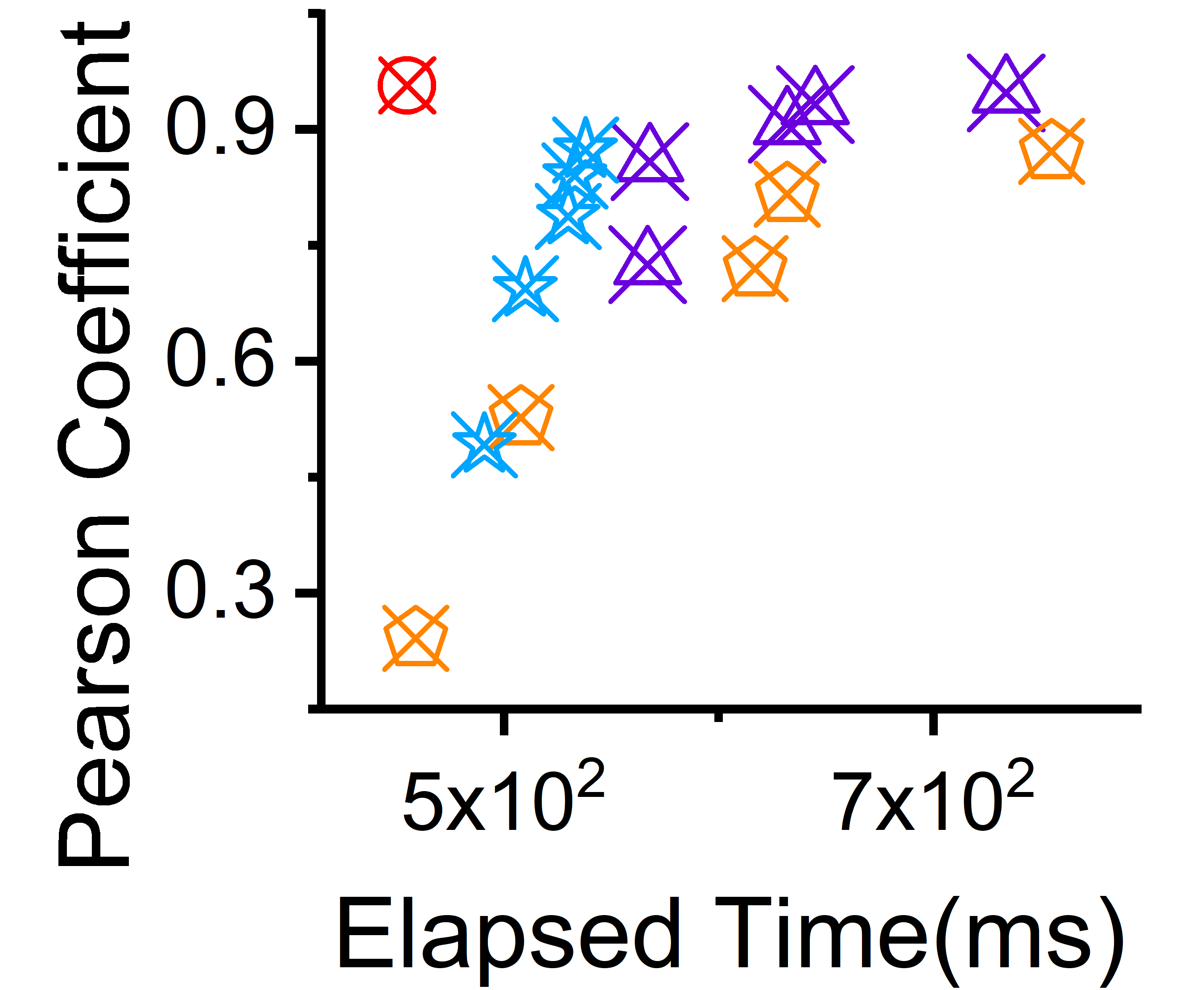}} 
    \\
    \vspace{-2mm}
    \caption[Speed and accuracy of DTC-FD.]{\figsummary{Speed and accuracy of DTC-FD.} 
        DTC-FD achieves the best trade-off between speed and accuracy compared to other algorithms for triangle counting in fully-dynamic graph streams.
    }
    \label{fig:fully_dynamic:multi_worker_time}
\end{figure*}

\begin{figure*}
    \centering
    \includegraphics[width=0.35\linewidth]{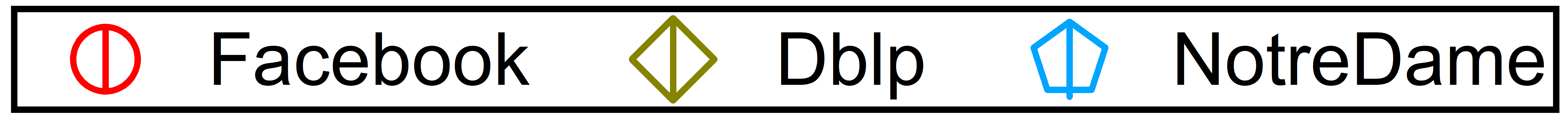}\\
    \vspace{0.5mm}
    \hspace{-4mm}
    \vspace{0.5mm}
    \subfigure[Global Error]{\includegraphics[width= 0.2\linewidth]{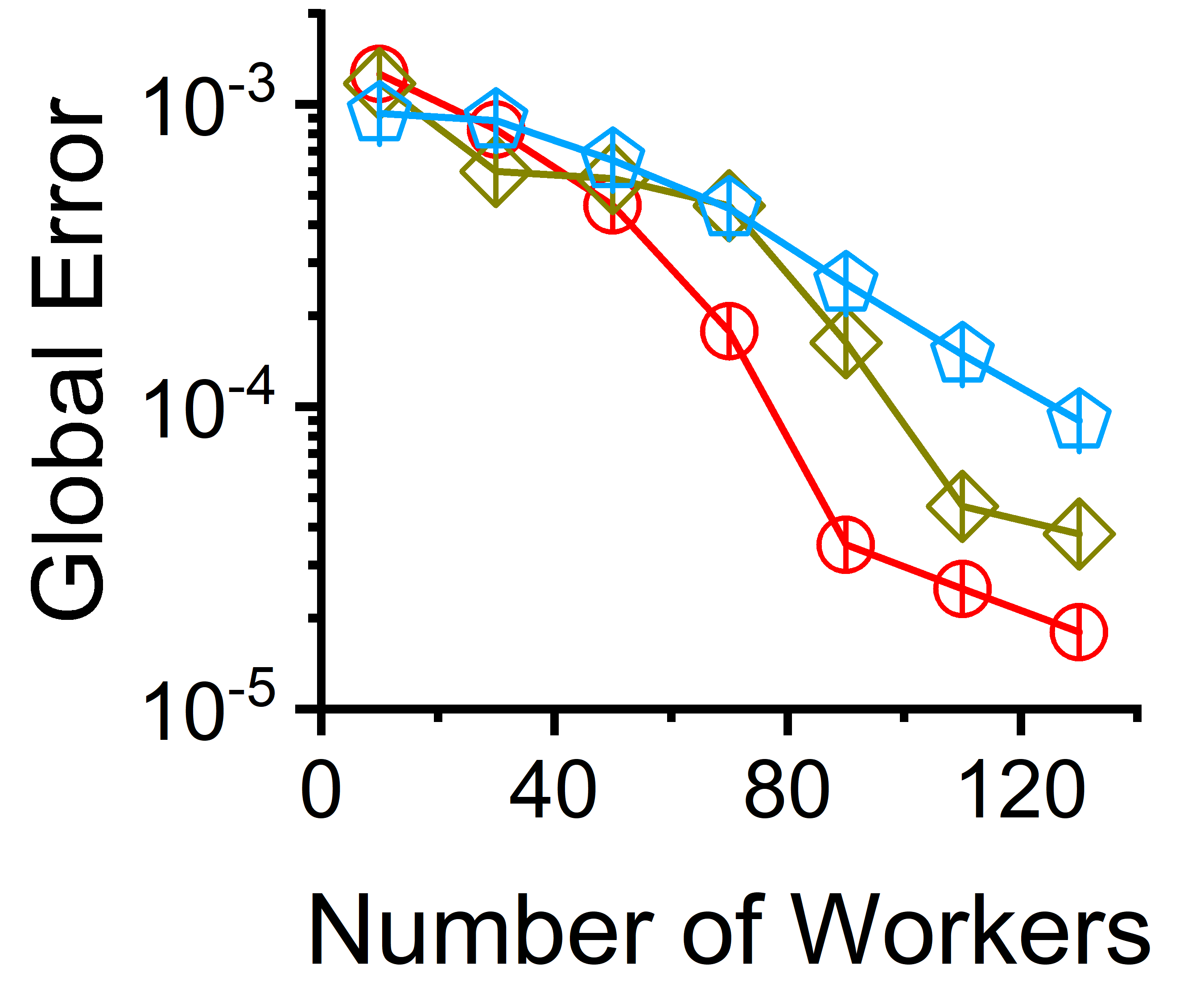}}
    \subfigure[Global Variance]{\includegraphics[width= 0.2\linewidth]{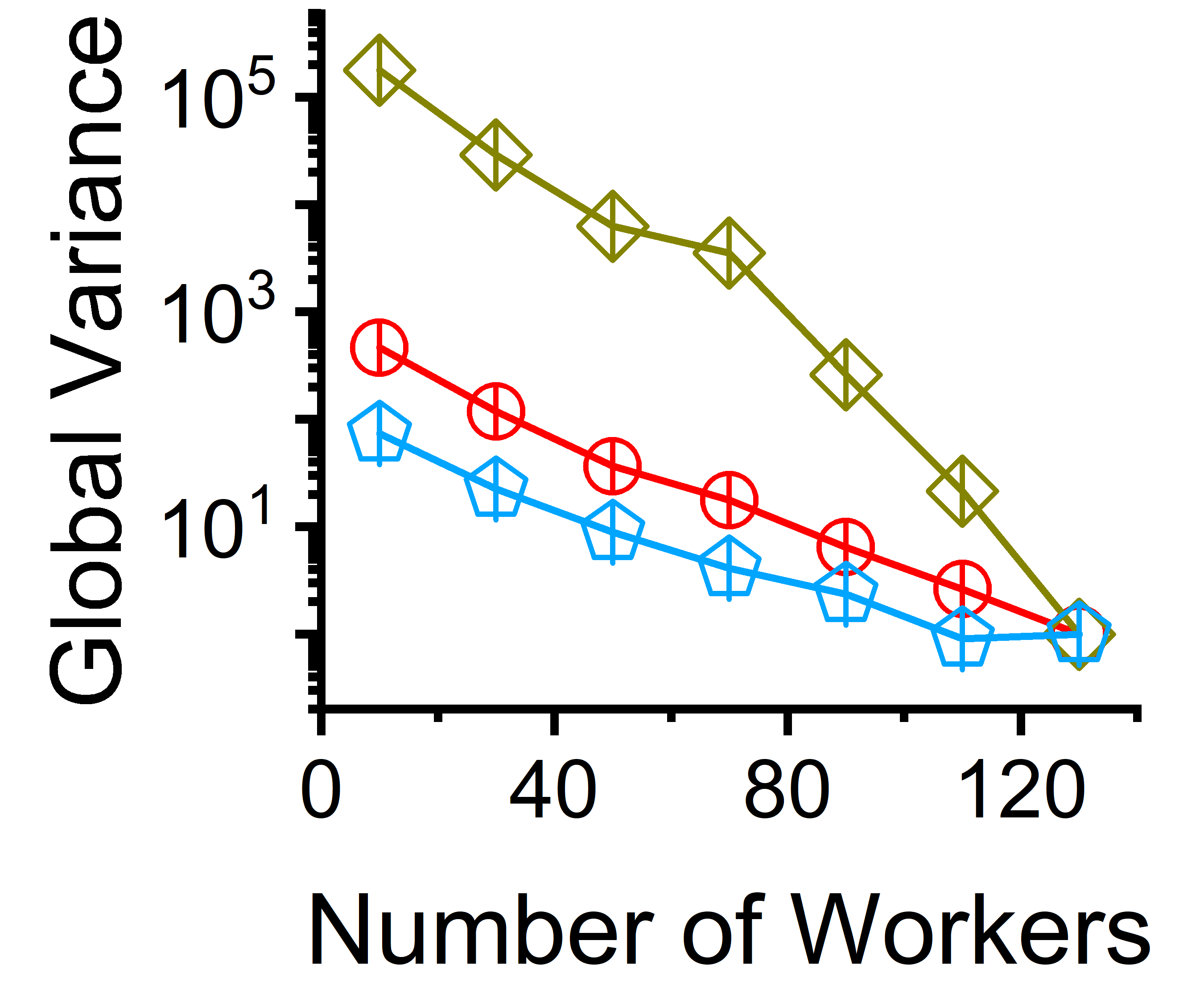}}
    \subfigure[Local Error]{\includegraphics[width= 0.2\linewidth]{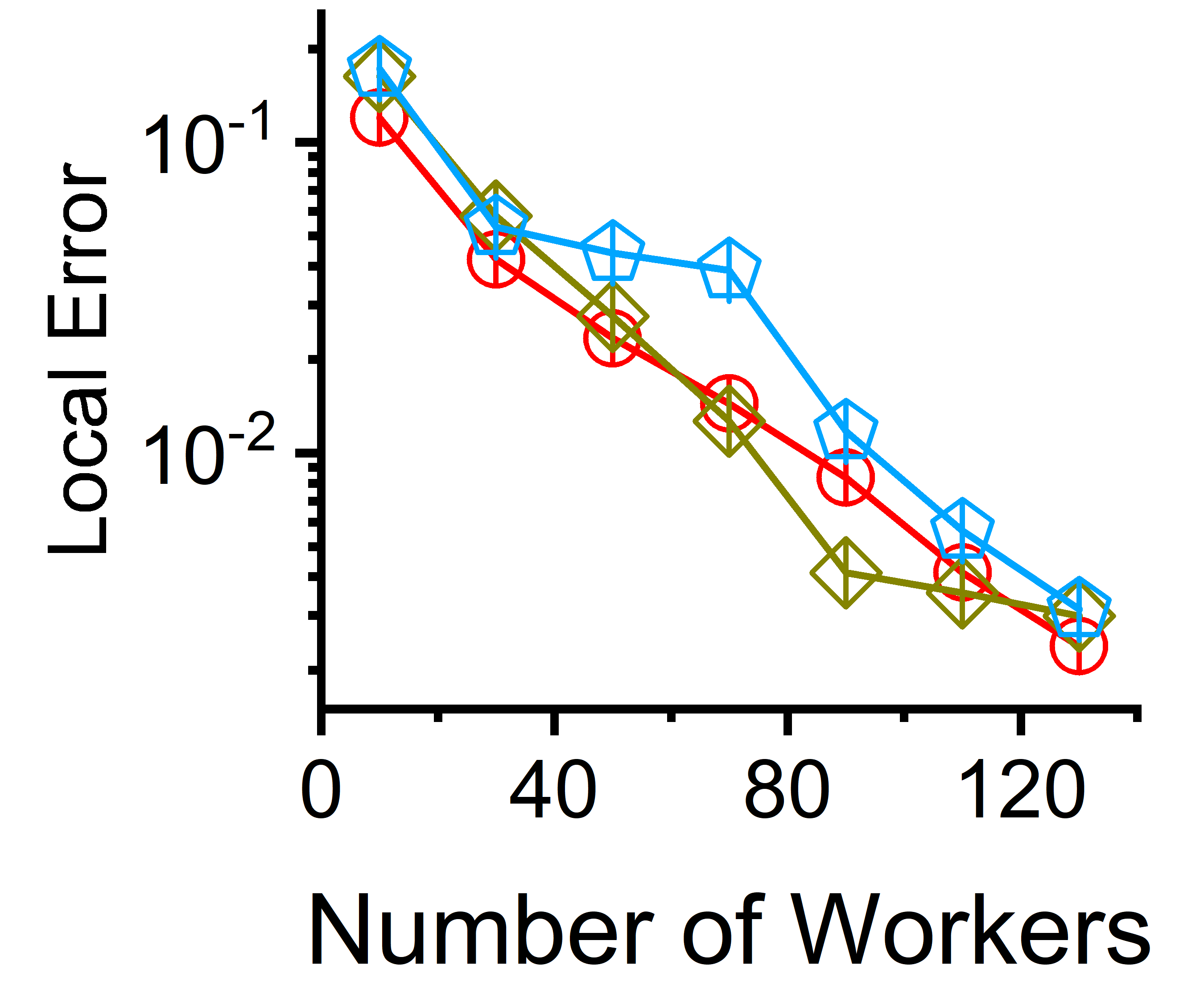}}
    \subfigure[Pearson Coefficient]{\includegraphics[width= 0.2\linewidth]{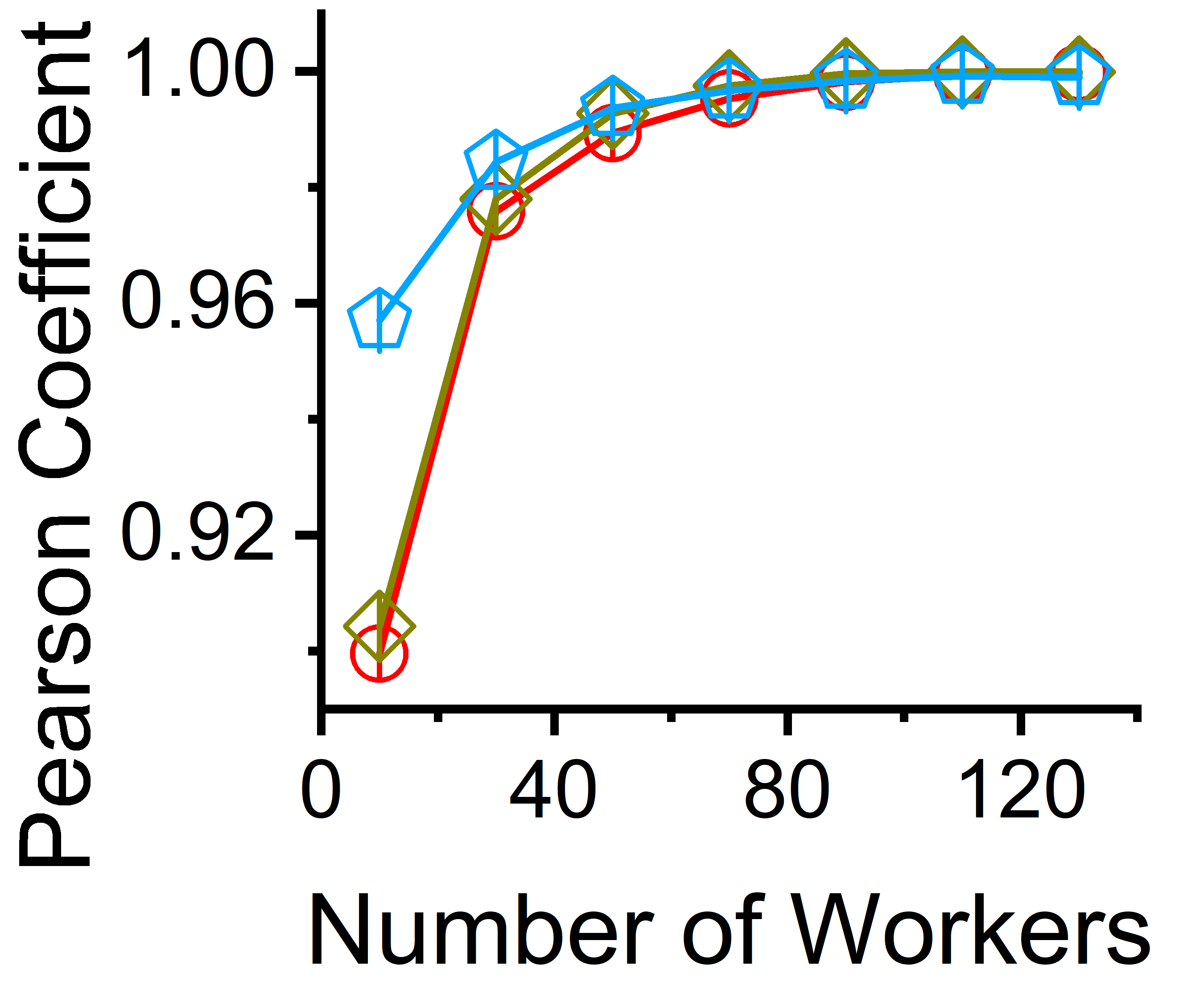}}
    \\
    \vspace{-2mm}
    \caption[Effects of Workers]{\figsummary{Effects of Workers.} 
        The estimation error significantly decreases as the number of workers increases, indicating the beneficial effects of employing more workers.
    }
    \label{fig:fully_dynamic:multi_worker}
\end{figure*}

\begin{figure}
    \centering
    \vspace{0.5mm}
    \hspace{-4mm}
    \vspace{0.5mm}
    \subfigure[Skitter]{\includegraphics[width= 0.35\linewidth]{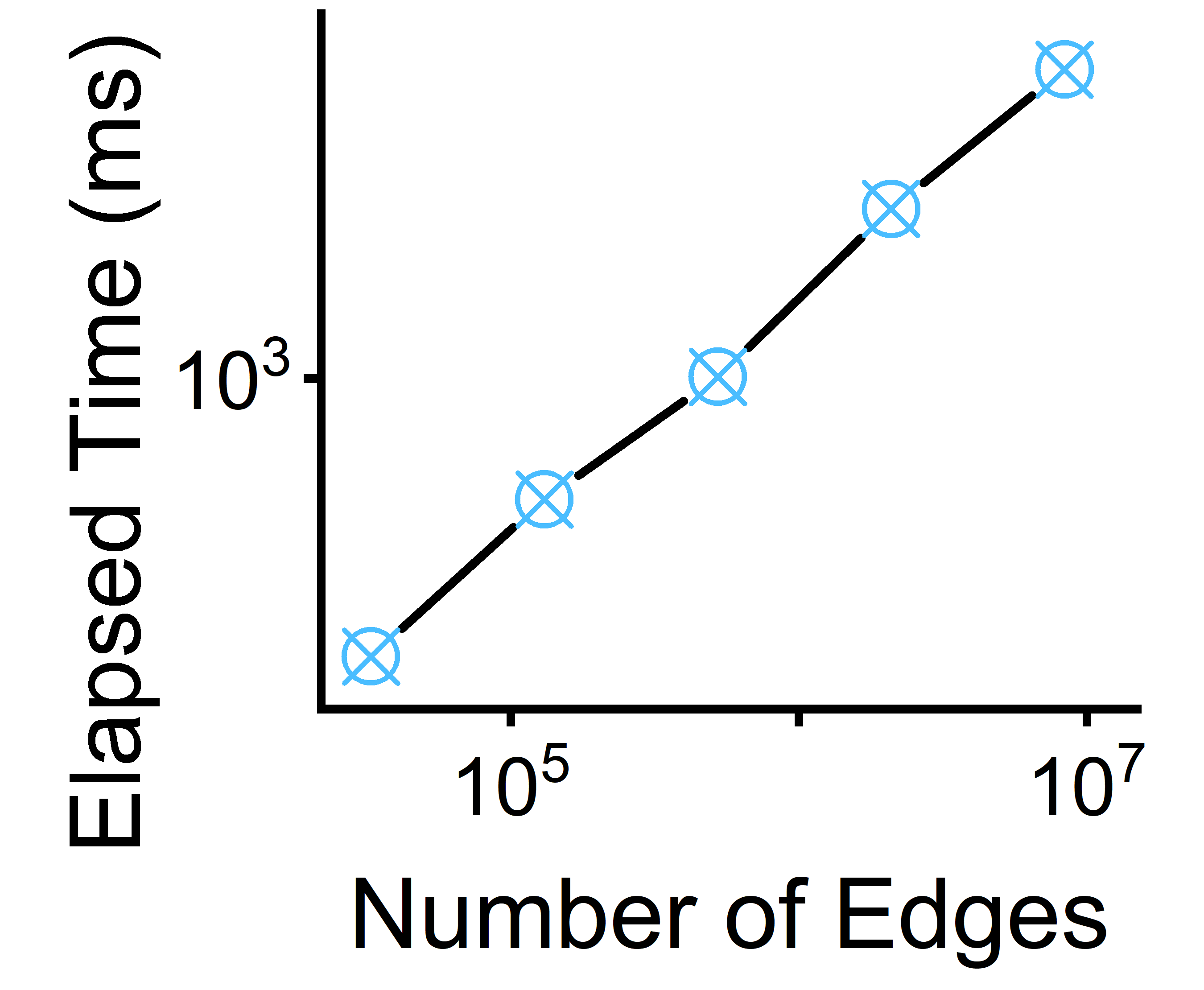}}
    \subfigure[LiveJournal]{\includegraphics[width= 0.35\linewidth]{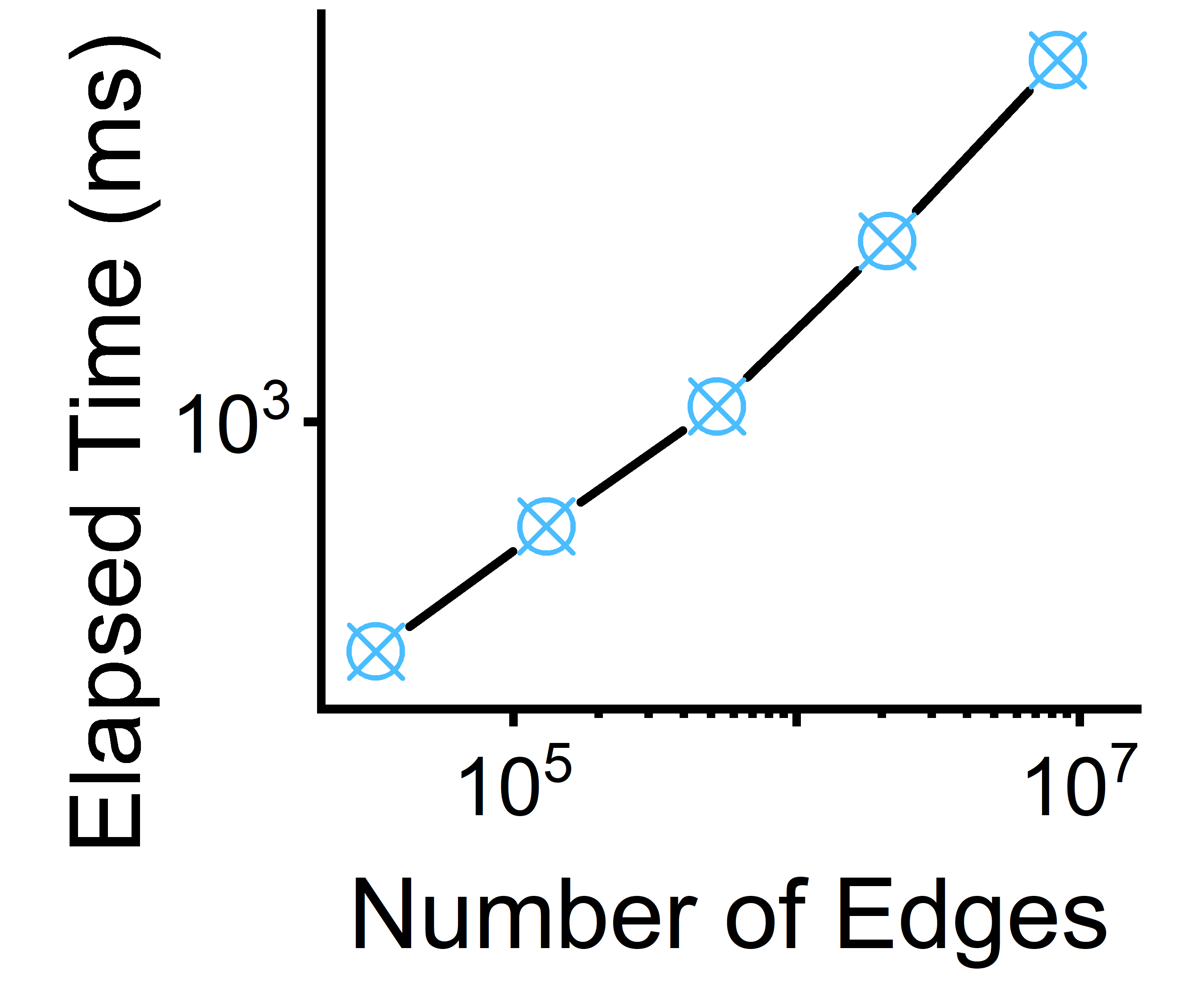}}
    \\
    \vspace{-2mm}
    \caption[Scalability of DTC-FD]{ \figsummary{Scalability of DTC-FD.} The time consumed by DTC-FD exhibits a linear relationship with the number of edges in fully dynamic streams, ensuring efficient processing as the edge count increases.
    }
    \label{fig:fully_dynamic:scalability}
\end{figure}

\subsection{DTC-FD for Fully Dynamic Streams}

Based on the real-world graph datasets in Table \ref{table:dataset}, we created fully dynamic graph streams by deleting $\delta$ of the edges and randomly locating them after these corresponding edge insertions. Here, we set $\delta$ to $20$\% unless otherwise stated. As far as we know, DTC-FD is the first distributed streaming algoritm to handle both edge insertions and deletions for global and local triangle counting in fully dynamic streams.

\subsubsection{Accuracy}

In Figure \ref{fig:fully_dynamic:compare}, we compard the accuracy between proposed DTC-FD and three state-of-the-art single-machine streaming algorithms, which all support fully dynamic graph streams. Here, we let the number of workers be 30, storage budget be 1\% of the size of graph streams, and each evaluation metric was conducted 100 times to get the average for all graph datasets. As shown in Figure \ref{fig:fully_dynamic:compare}, DTC-FD consistently yielded the best accuracy than others by utilizing multiple machines to store sampled edges for both global and local triangle counting. Specifically, DTC-FD was more accuracy than MASCOT-FD in terms of global error (up to 32.5$\times$) and local error (up to 19.3$\times$), and than ThinkDAcc in terms of global variance (up to 109.7$\times$) and pearson coefficient (up to 2.6$\times$), repectively.

\subsubsection{Speed}

In Figure \ref{fig:fully_dynamic:multi_worker_time}, we measured the accuracies and speeds between proposed DTC-FD and three state-of-the-art streaming algorithms supporting both edge insertions and deletions. Here, we let the number of workers be 10, storage budget be 1\% of the size of graph streams for DTC-FD, while 1\% to 5\% for other three algorithms, and conducted 100 times for each evaluation metric for graph datasets. As shown in Figure \ref{fig:fully_dynamic:multi_worker_time}, DTC-FD consistently yielded the best proformance for both speed and accuracy. Specifically, DTC-FD was up to 2.1$\times$ faster than ThinkDAcc with almost the same global error.

\subsubsection{Effects of Workers} 

In Figure \ref{fig:fully_dynamic:multi_worker}, we gave an insight into how the number of workers in distributed settings affects the accuracy for triangle counting in fully dynamic graph streams. Here, we let the storage budget be 1\% of the size of graph streams, and varied the number of workers from 10 to 130. And we conducted 1000 times for each evaluation metric to report their average. As illustrated in Figure \ref{fig:fully_dynamic:multi_worker}, DTC-FD yielded more accuracy, as more workers were added. Specifically, global error, global variance and local error dropped quickly with the increase of workers, and the pearson coefficient almost became 1 with about 130 workers.

\subsubsection{Scalability}

In Figure \ref{fig:fully_dynamic:scalability}, we verified how the number of edges of fully dynamic graph streams affects the speed for triangle counting in multiple machines. Here, we let the number of workers be 30, and storage budget be $10^4$ for graph datasets Skitter and LiveJournal. We measured the global error at the number of edges from $2^{15}$ to $2^{23}$ for Skitter, and from $2^{17}$ to $2^{25}$ for LiveJournal. As shown in Figure \ref{fig:fully_dynamic:scalability}, DTC-FD scaled linearly with the size of edges in fully dynamic graph streams.
\section{Conclusions}\label{sec-con}

In this article, we propose two novel distributed streaming algorithms for global and local triangle counting. These algorithms are single-pass and utilize standard reservoir sampling and random pairing techniques to ensure unbiased and accurate estimations. Our experimental results demonstrate the superiority of our proposed algorithms over other state-of-the-art approaches on real-world datasets. Specifically, DTC-AR achieves a better trade-off between accuracy and storage space by adaptive resampling without any prior knowledge about evolving graph streams. Compared to MASCOT and CoCoS, DTC-AR achieves up to 2029.4$\times$ and 27.1$\times$ smaller estimation errors, respectively. On the other hand, DTC-FD outperforms state-of-the-art baselines by providing up to 32.5$\times$ higher accuracy and being 2.1$\times$ faster. Additionally, DTC-FD exhibits linear scalability with the size of edges in fully dynamic streams. By leveraging these algorithms, we achieve significant improvements in accuracy, storage efficiency, and computational speed compared to existing approaches. 
Our future work will largely focus on estimating the count of general subgraphs (e.g. higher-order pattern) using multiple machines in fully dynamic graph streams. This will further enhance the capabilities of our algorithms and expand their applicability in analyzing complex network structures.

\section*{Acknowledgment}

This work was supported by National Key Research and Development Program (Grant No.2023YFB4502305), the Beijing Natural Science Foundation (4232036).

\bibliographystyle{IEEEtran}
\bibliography{references}

@article{jia2023persistent,
  title={Persistent graph stream summarization for real-time graph analytics},
  author={Jia, Yan and Gu, Zhaoquan and Jiang, Zhihao and Gao, Cuiyun and Yang, Jianye},
  journal={World Wide Web},
  volume={26},
  number={5},
  pages={2647--2667},
  year={2023},
  publisher={Springer}
}

@article{christopoulos2023local,
  title={Local Community Detection in Graph Streams with Anchors},
  author={Christopoulos, Konstantinos and Baltsou, Georgia and Tsichlas, Konstantinos},
  journal={Information},
  volume={14},
  number={6},
  pages={332},
  year={2023},
  publisher={MDPI}
}

@article{gemulla2008maintaining,
  title     = {Maintaining bounded-size sample synopses of evolving datasets},
  author    = {Gemulla, Rainer and Lehner, Wolfgang and Haas, Peter J},
  journal   = {The VLDB Journal},
  volume    = {17},
  number    = {2},
  pages     = {173--201},
  year      = {2008},
  publisher = {Springer}
}

@article{vitter1985random,
  title     = {Random sampling with a reservoir},
  author    = {Vitter, Jeffrey S},
  journal   = {ACM Transactions on Mathematical Software (TOMS)},
  volume    = {11},
  number    = {1},
  pages     = {37--57},
  year      = {1985},
  publisher = {ACM New York, NY, USA}
}

@article{shin2021cocos,
  title     = {Cocos: Fast and accurate distributed triangle counting in graph streams},
  author    = {Shin, Kijung and Lee, Euiwoong and Oh, Jinoh and Hammoud, Mohammad and Faloutsos, Christos},
  journal   = {ACM Transactions on Knowledge Discovery from Data (TKDD)},
  volume    = {15},
  number    = {3},
  pages     = {1--30},
  year      = {2021},
  publisher = {ACM New York, NY}
}

@inproceedings{xuantriangle,
  title={Triangle Counting by Adaptively Resampling over Evolving Graph Streams},
  author={Xuan, Wei and Cao, Huawei and Yan, Mingyu and Tang, Zhimin and Ye, Xiaochun and Fan, Dongrui},
  booktitle={33rd International Conference on Software Engineering \& Knowledge Engineering (SEKE)},
  pages={387--392},
  year={2021}
}

@article{stefani2017triest,
  title     = {Triest: Counting local and global triangles in fully dynamic streams with fixed memory size},
  author    = {Stefani, Lorenzo De and Epasto, Alessandro and Riondato, Matteo and Upfal, Eli},
  journal   = {ACM Transactions on Knowledge Discovery from Data (TKDD)},
  volume    = {11},
  number    = {4},
  pages     = {1--50},
  year      = {2017},
  publisher = {ACM New York, NY, USA}
}

@inproceedings{shin2018think,
  title        = {Think before you discard: Accurate triangle counting in graph streams with deletions},
  author       = {Shin, Kijung and Kim, Jisu and Hooi, Bryan and Faloutsos, Christos},
  booktitle    = {Joint European Conference on Machine Learning and Knowledge Discovery in Databases},
  pages        = {141--157},
  year         = {2018},
  organization = {Springer}
}

@inproceedings{leskovec2008microscopic,
  title     = {Microscopic evolution of social networks},
  author    = {Leskovec, Jure and Backstrom, Lars and Kumar, Ravi and Tomkins, Andrew},
  booktitle = {Proceedings of the 14th ACM SIGKDD international conference on Knowledge discovery and data mining},
  pages     = {462--470},
  year      = {2008}
}

@article{lim2018memory,
  title     = {Memory-efficient and accurate sampling for counting local triangles in graph streams: from simple to multigraphs},
  author    = {Lim, Yongsub and Jung, Minsoo and Kang, U},
  journal   = {ACM Transactions on Knowledge Discovery from Data (TKDD)},
  volume    = {12},
  number    = {1},
  pages     = {1--28},
  year      = {2018},
  publisher = {ACM New York, NY, USA}
}

@inproceedings{becchetti2008efficient,
  title     = {Efficient semi-streaming algorithms for local triangle counting in massive graphs},
  author    = {Becchetti, Luca and Boldi, Paolo and Castillo, Carlos and Gionis, Aristides},
  booktitle = {Proceedings of the 14th ACM SIGKDD international conference on Knowledge discovery and data mining},
  pages     = {16--24},
  year      = {2008}
}

@article{newman2003structure,
  title     = {The structure and function of complex networks},
  author    = {Newman, Mark EJ},
  journal   = {SIAM review},
  volume    = {45},
  number    = {2},
  pages     = {167--256},
  year      = {2003},
  publisher = {SIAM}
}

@article{assadi2023streaming,
  title={Streaming Algorithms and Lower Bounds for Estimating Correlation Clustering Cost},
  author={Assadi, Sepehr and Shah, Vihan and Wang, Chen},
  journal={Advances in Neural Information Processing Systems},
  volume={36},
  pages={75201--75213},
  year={2023}
}

@incollection{foucault2010friend,
  title     = {Is a friend a friend? Investigating the structure of friendship networks in virtual worlds},
  author    = {Foucault Welles, Brooke and Van Devender, Anne and Contractor, Noshir},
  booktitle = {CHI'10 Extended Abstracts on Human Factors in Computing Systems},
  pages     = {4027--4032},
  year      = {2010}
}

@inproceedings{ahmed2014graph,
  title     = {Graph sample and hold: A framework for big-graph analytics},
  author    = {Ahmed, Nesreen K and Duffield, Nick and Neville, Jennifer and Kompella, Ramana},
  booktitle = {Proceedings of the 20th ACM SIGKDD international conference on Knowledge discovery and data mining},
  pages     = {1446--1455},
  year      = {2014}
}

@article{ahmed2017sampling,
  author    = {Ahmed, Nesreen K. and Duffield, Nick and Willke, Theodore L. and Rossi, Ryan A.},
  title     = {On Sampling from Massive Graph Streams},
  journal   = {Proceedings of the VLDB Endowment},
  volume    = {10},
  number    = {11},
  year      = {2017},
  pages     = {1430--1441},
  publisher = {VLDB Endowment}
}

@article{pagh2012colorful,
  title     = {Colorful triangle counting and a mapreduce implementation},
  author    = {Pagh, Rasmus and Tsourakakis, Charalampos E},
  journal   = {Information Processing Letters},
  volume    = {112},
  number    = {7},
  pages     = {277--281},
  year      = {2012},
  publisher = {Elsevier}
}

@article{pavan2013counting,
  title     = {Counting and sampling triangles from a graph stream},
  author    = {Pavan, Aduri and Tangwongsan, Kanat and Tirthapura, Srikanta and Wu, Kun-Lung},
  journal   = {Proceedings of the VLDB Endowment},
  volume    = {6},
  number    = {14},
  pages     = {1870--1881},
  year      = {2013},
  publisher = {VLDB Endowment}
}

@inproceedings{kavassery2018improved,
  title        = {Improved triangle counting in graph streams: power of multi-sampling},
  author       = {Kavassery--Parakkat, Neeraj and Hanjani, Kiana Mousavi and Pavan, A},
  booktitle    = {2018 IEEE/ACM International Conference on Advances in Social Networks Analysis and Mining (ASONAM)},
  pages        = {33--40},
  year         = {2018},
  organization = {IEEE}
}

@inproceedings{shin2017wrs,
  title        = {Wrs: Waiting room sampling for accurate triangle counting in real graph streams},
  author       = {Shin, Kijung},
  booktitle    = {2017 IEEE International Conference on Data Mining (ICDM)},
  pages        = {1087--1092},
  year         = {2017},
  organization = {IEEE}
}

@article{shin2020fast,
  title     = {Fast, accurate and provable triangle counting in fully dynamic graph streams},
  author    = {Shin, Kijung and Oh, Sejoon and Kim, Jisu and Hooi, Bryan and Faloutsos, Christos},
  journal   = {ACM Transactions on Knowledge Discovery from Data (TKDD)},
  volume    = {14},
  number    = {2},
  pages     = {1--39},
  year      = {2020},
  publisher = {ACM New York, NY, USA}
}

@inproceedings{tsourakakis2009doulion,
  title     = {Doulion: counting triangles in massive graphs with a coin},
  author    = {Tsourakakis, Charalampos E and Kang, U and Miller, Gary L and Faloutsos, Christos},
  booktitle = {Proceedings of the 15th ACM SIGKDD international conference on Knowledge discovery and data mining},
  pages     = {837--846},
  year      = {2009}
}

@inproceedings{mcgregor2016better,
  title     = {Better algorithms for counting triangles in data streams},
  author    = {McGregor, Andrew and Vorotnikova, Sofya and Vu, Hoa T},
  booktitle = {Proceedings of the 35th ACM SIGMOD-SIGACT-SIGAI Symposium on Principles of Database Systems},
  pages     = {401--411},
  year      = {2016}
}

@inproceedings{shin2018tri,
  title        = {Tri-fly: Distributed estimation of global and local triangle counts in graph streams},
  author       = {Shin, Kijung and Hammoud, Mohammad and Lee, Euiwoong and Oh, Jinoh and Faloutsos, Christos},
  booktitle    = {Pacific-Asia Conference on Knowledge Discovery and Data Mining},
  pages        = {651--663},
  year         = {2018},
  organization = {Springer}
}

@inproceedings{yu2019distributed,
  title        = {Distributed triangle counting algorithms in simple graph stream},
  author       = {Yu, Mengdi and Song, Chao and Gu, Jiqing and Liu, Ming},
  booktitle    = {2019 IEEE 25th International Conference on Parallel and Distributed Systems (ICPADS)},
  pages        = {294--301},
  year         = {2019},
  organization = {IEEE}
}

@article{yang2022distributed,
  title     = {Distributed Triangle Approximately Counting Algorithms in Simple Graph Stream},
  author    = {Yang, Xu and Song, Chao and Yu, Mengdi and Gu, Jiqing and Liu, Ming},
  journal   = {ACM Transactions on Knowledge Discovery from Data (TKDD)},
  volume    = {16},
  number    = {4},
  pages     = {1--43},
  year      = {2022},
  publisher = {ACM New York, NY}
}

@inproceedings{wang2019rept,
  title        = {REPT: A streaming algorithm of approximating global and local triangle counts in parallel},
  author       = {Wang, Pinghui and Jia, Peng and Qi, Yiyan and Sun, Yu and Tao, Jing and Guan, Xiaohong},
  booktitle    = {2019 IEEE 35th International Conference on Data Engineering (ICDE)},
  pages        = {758--769},
  year         = {2019},
  organization = {IEEE}
}

@inproceedings{rahman2013approximate,
  title        = {Approximate triangle counting algorithms on multi-cores},
  author       = {Rahman, Mahmudur and Al Hasan, Mohammad},
  booktitle    = {2013 IEEE International Conference on Big Data},
  pages        = {127--133},
  year         = {2013},
  organization = {IEEE}
}

@inproceedings{shun2015multicore,
  title        = {Multicore triangle computations without tuning},
  author       = {Shun, Julian and Tangwongsan, Kanat},
  booktitle    = {2015 IEEE 31st International Conference on Data Engineering},
  pages        = {149--160},
  year         = {2015},
  organization = {IEEE}
}

@inproceedings{hu2018tricore,
  title        = {Tricore: Parallel triangle counting on gpus},
  author       = {Hu, Yang and Liu, Hang and Huang, H Howie},
  booktitle    = {SC18: International Conference for High Performance Computing, Networking, Storage and Analysis},
  pages        = {171--182},
  year         = {2018},
  organization = {IEEE}
}

@inproceedings{hu2021accelerating,
  title     = {Accelerating triangle counting on gpu},
  author    = {Hu, Lin and Zou, Lei and Liu, Yu},
  booktitle = {Proceedings of the 2021 International Conference on Management of Data},
  pages     = {736--748},
  year      = {2021}
}

@inproceedings{park2013efficient,
  title     = {An efficient mapreduce algorithm for counting triangles in a very large graph},
  author    = {Park, Ha-Myung and Chung, Chin-Wan},
  booktitle = {Proceedings of the 22nd ACM international conference on Information \& Knowledge Management},
  pages     = {539--548},
  year      = {2013}
}

@inproceedings{arifuzzaman2015space,
  title        = {A space-efficient parallel algorithm for counting exact triangles in massive networks},
  author       = {Arifuzzaman, Shaikh and Khan, Maleq and Marathe, Madhav},
  booktitle    = {2015 IEEE 17th International Conference on High Performance Computing and Communications, 2015 IEEE 7th International Symposium on Cyberspace Safety and Security, and 2015 IEEE 12th International Conference on Embedded Software and Systems},
  pages        = {527--534},
  year         = {2015},
  organization = {IEEE}
}

@article{bisson2017high,
  title     = {High performance exact triangle counting on gpus},
  author    = {Bisson, Mauro and Fatica, Massimiliano},
  journal   = {IEEE Transactions on Parallel and Distributed Systems},
  volume    = {28},
  number    = {12},
  pages     = {3501--3510},
  year      = {2017},
  publisher = {IEEE}
}

@article{horvitz1952generalization,
  title     = {A generalization of sampling without replacement from a finite universe},
  author    = {Horvitz, Daniel G and Thompson, Donovan J},
  journal   = {Journal of the American statistical Association},
  volume    = {47},
  number    = {260},
  pages     = {663--685},
  year      = {1952},
  publisher = {Taylor \& Francis}
}

@inproceedings{lim2015mascot,
  title     = {Mascot: Memory-efficient and accurate sampling for counting local triangles in graph streams},
  author    = {Lim, Yongsub and Kang, U},
  booktitle = {Proceedings of the 21th ACM SIGKDD international conference on knowledge discovery and data mining},
  pages     = {685--694},
  year      = {2015}
}

@inproceedings{kutzkov2014triangle,
  title        = {Triangle counting in dynamic graph streams},
  author       = {Kutzkov, Konstantin and Pagh, Rasmus},
  booktitle    = {Scandinavian Workshop on Algorithm Theory},
  pages        = {306--318},
  year         = {2014},
  organization = {Springer}
}

@article{jha2015space,
  title     = {A space-efficient streaming algorithm for estimating transitivity and triangle counts using the birthday paradox},
  author    = {Jha, Madhav and Seshadhri, C and Pinar, Ali},
  journal   = {ACM Transactions on Knowledge Discovery from Data (TKDD)},
  volume    = {9},
  number    = {3},
  pages     = {1--21},
  year      = {2015},
  publisher = {ACM New York, NY, USA}
}

@inproceedings{tsourakakis2008fast,
  title        = {Fast counting of triangles in large real networks without counting: Algorithms and laws},
  author       = {Tsourakakis, Charalampos E},
  booktitle    = {2008 Eighth IEEE International Conference on Data Mining},
  pages        = {608--617},
  year         = {2008},
  organization = {IEEE}
}

@article{kolountzakis2012efficient,
  title     = {Efficient triangle counting in large graphs via degree-based vertex partitioning},
  author    = {Kolountzakis, Mihail N and Miller, Gary L and Peng, Richard and Tsourakakis, Charalampos E},
  journal   = {Internet Mathematics},
  volume    = {8},
  number    = {1-2},
  pages     = {161--185},
  year      = {2012},
  publisher = {Taylor \& Francis}
}

@article{pavan2013parallel,
  title   = {Parallel and distributed triangle counting on graph streams},
  author  = {Pavan, Aduri and Tangwongan, Kanat and Tirthapura, Srikanta},
  journal = {Technical report, IBM, Tech. Rep.},
  year    = {2013}
}

@article{gehrke2003overview,
  title     = {Overview of the 2003 KDD Cup},
  author    = {Gehrke, Johannes and Ginsparg, Paul and Kleinberg, Jon},
  journal   = {Acm Sigkdd Explorations Newsletter},
  volume    = {5},
  number    = {2},
  pages     = {149--151},
  year      = {2003},
  publisher = {ACM New York, NY, USA}
}

@inproceedings{viswanath2009evolution,
  title     = {On the evolution of user interaction in facebook},
  author    = {Viswanath, Bimal and Mislove, Alan and Cha, Meeyoung and Gummadi, Krishna P},
  booktitle = {Proceedings of the 2nd ACM workshop on Online social networks},
  pages     = {37--42},
  year      = {2009}
}

@misc{snapnets,
  author       = {Jure Leskovec and Andrej Krevl},
  title        = {{SNAP Datasets}: {Stanford} Large Network Dataset Collection},
  howpublished = {\url{http://snap.stanford.edu/data}},
  month        = jun,
  year         = 2014
}

@inproceedings{mislove2007measurement,
  title     = {Measurement and analysis of online social networks},
  author    = {Mislove, Alan and Marcon, Massimiliano and Gummadi, Krishna P and Druschel, Peter and Bhattacharjee, Bobby},
  booktitle = {Proceedings of the 7th ACM SIGCOMM conference on Internet measurement},
  pages     = {29--42},
  year      = {2007}
}

\end{document}